\documentclass[twocolumn,natbib]{article}

\usepackage{amssymb}
\usepackage{bbm}
\usepackage{natbib}
\usepackage{psfrag}
\usepackage{graphicx}
\usepackage{subfigure}
\usepackage{mathtools}
\usepackage{dsfont, multicol}

\usepackage{color}

\usepackage{algorithmic}
\algsetup{linenosize=\small}
\usepackage{algorithm}

\begin{document}

% \mathtoolsset{showonlyrefs,showmanualtags}
\begin{multicols}{2}
\title{\bf Fast Approximate Bayesian Computation for discretely observed Markov models using a factorised posterior distribution}
\author{
    Simon R. White\\
    MRC Biostatistics Unit, Cambridge \\
    {\tt simon.white@mrc-bsu.cam.ac.uk}
     \and
    Theodore Kypraios\\
    University of Nottingham\\
    {\tt theodore.kypraios@nottingham.ac.uk}
   \and
   Simon P. Preston\\
   University of Nottingham\\
  {\tt Simon.Preston@nottingham.ac.uk}    
}
\date{May, 25, 2013}
\maketitle
\end{multicols}
\maketitle

%%%%
%%%%
%%%%
%%%% Brief Report max 3500 words
%%%%
%%%%
%%%%
%%%%
%%%%
%%%%
%%%%
%
% Notes are placed on post submission formatting
%
%
%

\begin{abstract} Many modern statistical applications involve inference for complicated
  stochastic models for which the likelihood function is difficult or even impossible to
  calculate, and hence conventional likelihood-based inferential techniques cannot be
  used.  In such settings, Bayesian inference can be performed using Approximate Bayesian
  Computation (ABC).  However, in spite of many recent developments to ABC methodology, in
  many applications the computational cost of ABC necessitates the choice of summary
  statistics and tolerances that can potentially severely bias the estimate of the
  posterior.

  We propose a new ``piecewise'' ABC approach suitable for discretely observed Markov
  models that involves writing the posterior density of the parameters as a product of
  factors, each a function of only a subset of the data, and then using ABC within each
  factor.  The approach has the advantage of side-stepping the need to choose a summary
  statistic and it enables a stringent tolerance to be set, making the posterior ``less
  approximate''.  We investigate two methods for estimating the posterior density based on
  ABC samples for each of the factors: the first is to use a Gaussian approximation for
  each factor, and the second is to use a kernel density estimate.  Both methods have
  their merits. The Gaussian approximation is simple, fast, and probably adequate for many
  applications.  On the other hand, using instead a kernel density estimate has the
  benefit of consistently estimating the true piecewise ABC posterior as the number of ABC
  samples tends to infinity.   We illustrate the piecewise ABC approach with four
  examples; in each case, the approach offers fast and accurate inference.

%  \PACS{05.45.Tp \and 05.10.-a \and 05.40.-a \and 02.50.-r \and 02.70.-c}
%05.45.Tp   (Time series in nonlinear dynamic - on several of the reference papers from
%PRE)
%% Statistical Physics
%02.50.-r   Probability theory, stochastic processes, and statistics 05.40.-a Fluctuation
%phenomena, random processes, noise, Brownian (also listed with stochastic processes)
%% Computational Physics
%02.70.-c   Computational techniques; simulations 05.10.-a   Computational methods in
%statistical physics and nonlinear dynamics
\end{abstract}

\section{Introduction}
\label{sec:outline}

Stochastic models are commonly used to model processes in the physical sciences
\citep{Wil06,van2007stochastic}. For many such models the likelihood is difficult or
costly to compute making it infeasible to use conventional inference techniques such as
maximum likelihood estimation.  However, provided it is possible to simulate from a model,
then ``implicit'' methods such as Approximate Bayesian Computation (ABC) methods enable 
inference without having to calculate the likelihood. These methods were
originally developed for applications in population genetics
\citep{pritchard:population_growth_of_Y_chormosomes:99} and human demographics
\citep{beaumont_zhang_balding:abc_pop_genetics:02}, but are now being used in a wide range
of fields including epidemiology
\citep{McKinley+Cook+Deardon:infer_epidemic_models_wo_likelihoods:09}, evolution of
species \citep{toni_welch_strelkowa_ipsen_stumpf:abc_dynamical_systems:09}, finance
\citep{Peters2010}, and evolution of pathogens
\citep{Wilson_Gabriel_Leatherbarrow_Cheesbrough_Gee_Bolton_Fox_Hart_Diggle_Fearnhead_2009},
to name a few.

Intuitively, ABC methods involve simulating data from the model using various parameter
values and making inference based on which parameter values produced realisations that are
``close'' to the observed data.  Let the data $x = \left( x_1, \ldots, x_n \right) \equiv
\left( x(t_1), \ldots, x(t_n) \right)$ be a vector comprising observations of a possibly
vector state variable $X(t)$ at time points $t_1, \ldots, t_n$.  We assume that the data
arise from a Markov stochastic model (which encompasses IID data as a special case)
parameterised by the vector $\theta$, which is the target of inference, and we denote by
$\pi(x | \theta)$ the probability density of the data given a specific value of $\theta.$
Prior beliefs about $\theta$ are expressed via a density denoted $\pi(\theta)$.  Algorithm 1 generates {\em exact} samples from the Bayesian posterior density
$\pi(\theta \vert x)$ which is proportional to $\pi(x \vert \theta) \pi(\theta)$:
\begin{algorithm}[H]
{\bf 1}: Sample $\theta^*$ from $\pi(\theta)$. \\
{\bf 2}: Simulate dataset ${x^*}$ from the model using parameters $\theta^*$. \\
{\bf 3}: Accept $\theta^*$ if $x^* = x$, otherwise reject. \\
{\bf 4}: Repeat.
\caption{\newline Exact Bayesian Computation (EBC)}
\label{alg:EBC}
\end{algorithm}

This algorithm is only of practical use if ${X(t)}$ is discrete, else
the acceptance probability in Step 3 is zero. For continuous
distributions, or discrete ones in which the acceptance probability in
step 3 is unacceptably low,
\citet{pritchard:population_growth_of_Y_chormosomes:99} suggested the
following algorithm:
\begin{algorithm}[H]
  \vspace{0.1cm}
As Algorithm 1, but with step 3 replaced by:\\
{\bf 3}': Accept $\theta^*$ if $d\big(s(x),s(x^*)\big) \leq \varepsilon$, otherwise
reject.
\caption{\newline Approximate Bayesian Computation (ABC)}
\label{alg:ABC}
\end{algorithm}
\noindent where $d(\cdot,\cdot)$ is a distance function, usually taken to be the
$L^2$-norm of the difference between its arguments; $s(\cdot)$ is a function of the data;
and $\varepsilon$ is a tolerance.  Note that $s(\cdot)$ can be the identity function but
in practice, to give tolerable acceptance rate, it is usually taken to be a
lower-dimensional vector 
comprising summary statistics that characterise key aspects of the data.

The output of the ABC algorithm is a sample from the ABC posterior density
%\begin{equation}
$\tilde{\pi}(\theta|x) =  \pi(\theta\vert x, d \big(s(x),s(x^*)\big)\leq\varepsilon). $
Provided
$s(\cdot)$ is sufficient for $\theta$, then the ABC posterior density converges to
$\pi(\theta \vert x)$ as $\varepsilon \rightarrow 0$ \citep{Marin2011}.  However, in practise it
is rarely possible to use an $s(\cdot)$ which is sufficient, or to take $\varepsilon$
especially small (or zero). Hence ABC requires a careful choice of $s(\cdot)$ and
$\varepsilon$ to make the acceptance rate tolerably large, at the same time as
trying not to make the ABC posterior too different from the true posterior, $\pi(\theta
\vert x)$.  In other words, there is a balance which involves trading off Monte
Carlo error with ``ABC error'' owing to the choice of $s(\cdot)$ and tolerance
$\varepsilon$.

Over the last decade, a wide range of extensions to the original ABC
algorithm have been developed, including Markov Chain Monte Carlo
(MCMC) \citep{MarMolPlagTav03} and Sequential Monte Carlo (SMC)
\citep{toni_welch_strelkowa_ipsen_stumpf:abc_dynamical_systems:09}
implementations, the incorporation of auxiliary regression models
\citep{beaumont_zhang_balding:abc_pop_genetics:02,BlumFran10},
%variational approximation \citep{Barthelme+Chopin:EP-ABC:11}, 
and (semi-)automatic choice
of summary statistics \citep{Fearnhead+Prangle}; see \citet{Marin2011}
for a review.  In all of these ABC variants, however, 
computational cost remains a central issue, since it is the computational cost that
determines the balance that can be made between controlling Monte Carlo error and
controlling bias arising from using summary statistics and/or non-zero tolerance.

In this paper we propose a novel algorithm called \emph {piecewise ABC} (PW-ABC), the aim
of which is to substantially reduce the computational cost of ABC.  The algorithm is
applicable to a particular (but fairly broad) class of models, namely those with the
Markov property and for which the state variable is observable at discrete time points.
The algorithm is based on a factorisation of the posterior density such that each factor
corresponds to only a subset of the data. The idea is to apply Algorithm 2 for each
factor (a task which is computationally very cheap), to compute the density estimates for
each factor, and then to estimate the full posterior density as the product of these factors.
Taking advantage of the factorisation lowers the computational burden of ABC such that the
choice of summary statistic and tolerance---and the accompanying biases---can potentially
be avoided completely.

In the following section we describe PW-ABC in more detail.  The main practical issue of
the method is how to use the ABC samples from each posterior factor to estimate the full
posterior density.  We discuss two approaches to estimating the relevant densities and
products of densities, then we apply PW-ABC, using both approaches, to four examples: a
toy illustrative example of inferring the probability of success in a binomial experiment,
a stochastic-differential-equation model, an autoregressive time-series model, and a dynamical
predator--prey model. We conclude with a discussion of the strengths and limitations of
PW-ABC, and of potential further generalisations.

\section{Piece-wise ABC (PW-ABC)} \label{sec:methods}

% In this section we describe the PW-ABC algorithm.  Note that there is also a piecewise
% analogue of EBC which is covered as a special case of PW-ABC, namely with $\varepsilon$
% taken to be zero and $s(\cdot)$ taken as the identity function.

Our starting point is to use the Markov property to write the likelihood as
\begin{align}
\pi(x \vert \theta) &= \left(\prod_{i=2}^n \pi (x_i \vert x_{i-1}, 
\ldots,x_1, \theta) \right) \pi(x_1 \vert \theta) \notag\\
&= \left(\prod_{i=2}^n \pi (x_i \vert x_{i-1},\theta)\right) \pi(x_1\vert \theta).
\label{eqn:factorised:likelihood}
\end{align}
The likelihood contribution of the first observation $x_1$ is asymptotically irrelevant as 
the number of observations, $n$, increases and, henceforth, to keep the presentation simple, we 
ignore the term $\pi(x_1 \vert \theta)$ in (\ref{eqn:factorised:likelihood}).
Accounting for this, and by using multiple applications of Bayes' theorem, the posterior density can be written in the following factorised form,
\begin{align}
\pi(\theta \vert x) &\propto \pi(x \vert \theta) \pi(\theta) \notag\\
&
= \left( \prod_{i=2}^n \frac{\pi( x_i \vert x_{i-1},\theta) \pi(\theta)}{\pi(\theta)} \right) %\pi(x_1\vert \theta) 
\pi(\theta) \notag\\
&\propto \pi(\theta)^{(2-n)} %\pi(\theta \vert x_1 )
% \left( \prod_{i=2}^n \pi( x_i \vert x_{i-1},\theta) \pi(\theta) \right).
\left( \prod_{i=2}^n \varphi_i(\theta) \right),
\label{eqn:factorised:posterior}
\end{align}
where
\begin{align*}
\varphi_i(\theta) &=  c_i^{-1} \pi( x_i \vert x_{i-1},\theta) \pi(\theta)\\
c_i &= \int \pi( x_i \vert x_{i-1},\theta) \pi(\theta) \mathrm{d} \theta.
\end{align*}
Essentially, in (\ref{eqn:factorised:posterior}) the posterior
density, $\pi(\theta \vert x)$, of $\theta$ given the
full data $x$ has been decomposed into a product
involving densities $\varphi_i(\theta)$, each of which depends only on
a pair of data points, $\{ x_{i-1}, x_{i} \}$.

The key idea now is to use ABC to draw approximate 
%(or exact, if $s(\cdot)=\mathrm{Identity}(\cdot)$
%and $\varepsilon=0$) 
samples from each of the densities $\varphi_i(\theta)$.  Applying
Algorithm~\ref{alg:ABC} involves (i) drawing $\theta^*$ from $\pi(\theta)$, (ii)
simulating $x_i^*|x_{i-1},\theta^*$, and (iii) accepting $\theta^*$ if
\mbox{$d\big(s(x_i),s(x_{i}^*)\big) \leq \varepsilon$}. 
We use $\tilde{\varphi}_i(\theta)$ to denote the implied ABC density from which these
samples are drawn (with $\tilde{\varphi}_i(\theta) = \varphi_i(\theta)$ if $s(\cdot)=\mathrm{Identity}(\cdot)$
and $\varepsilon=0$).  By repeating (i)---(iii) we generate samples
of, say, $m$ draws, $\theta^*_{i(1)}, \dots, \theta^*_{i(m)}$, from
each $\tilde{\varphi}_i(\theta)$.  Now, suppose that $\hat{\varphi}_i(\theta)$ is an estimate,
based on the sample \linebreak $\theta^*_{i(1)}, \dots, \theta^*_{i(m)}$, of the density
$\tilde{\varphi}_i(\theta)$ (and hence of the density $\varphi_i(\theta)$).  Then the
posterior density \eqref{eqn:factorised:posterior} can be estimated by
\begin{equation}
\hat{\pi}(\theta \vert x) = g(\theta) \Big\slash \!\!\int \! g(\theta) \, \mathrm{d}
\theta, \label{eqn:approx:posterior}
\end{equation}
where
\begin{equation}
g(\theta) = 
\pi(\theta)^{(2-n)} %\pi(\theta \vert x_1 )
\left( \prod_{i=2}^n \hat{\varphi}_i(\theta) \right).
\label{eqn:g:in:expression:for:posterior}
\end{equation}
The steps of PW-ABC are summarised in Algorithm 3.

\begin{algorithm}[h!]
\begin{algorithmic}
\FOR{$i=2$ to $n$}
\STATE {\small a:} Apply the ABC Algorithm to draw $m$ 
approximate (or exact, if $s(\cdot)=\mathrm{Identity}(\cdot)$ and $\varepsilon=0$)
samples, $\theta^*_{i(1)}, \dots, \theta^*_{i(m)}$, from $\varphi_i(\theta)$;
\STATE {\small b:} Using the samples $\theta^*_{i(1)}, \dots, \theta^*_{i(m)}$ and either
\eqref{eqn:gaussian:approx:to:component:density} or \eqref{eqn:KDE}, calculate a density 
estimate, $\hat{\varphi}_i(\theta)$, of $\tilde{\varphi}_i(\theta)$, the implied ABC
density.
\ENDFOR
\STATE Substitute the density estimates $\hat{\varphi}_i(\theta)$ into \eqref{eqn:approx:posterior} to 
calculate an estimate, $\hat{\pi}(\theta \vert x)$, of $\pi(\theta \vert
x).$
\end{algorithmic}
\caption{Piece-Wise Approximate Bayesian Computation (PW-ABC)}
\label{alg:PW-ABC}
\end{algorithm}

The rationale of the piecewise approach is to reduce the dimension for ABC, replacing a
high-dimensional problem with multiple low-dimensional ones.  In standard ABC the 
summary statistic, $s(\cdot)$, is the tool used to reduce the dimension, but 
in PW-ABC, with dimension already reduced by the factorisation in
(\ref{eqn:factorised:posterior}), we can take $s(\cdot)=\mathrm{Identity}(\cdot)$ and
typically use a much smaller $\varepsilon$.

The question remains of how to calculate the density estimates,
$\hat{\varphi}_i(\theta)$.  Below we discuss two approaches: (i) using a Gaussian
approximation, and (ii) using a kernel density estimate.  Henceforth, quantities based
on (i) are denoted by superscript g, and those based on (ii) are denoted by superscript
k.  In both cases we discuss the behaviour of the estimators in the asymptotic regime in
which the number of observations, $n$, is kept fixed while the size of each ABC sample
increases, $m$ $\rightarrow \infty$.

\subsection{Gaussian approximation for $\hat{\varphi}_i(\theta)$}
\label{sec:gauss-appr-hatv}

Denote the $d$-dimensional multivariate Gaussian density with mean, $\mu$, and covariance, $\Sigma$, by
\begin{align}
  K(\theta; \mu, \Sigma) &= (2\pi)^{-d/2} \, (\det \Sigma)^{-1/2} \times \nonumber \\
  &\quad\exp\left(
  -\tfrac{1}{2}{\left(\theta-\mu\right)^T
  \Sigma^{-1}\left(\theta-\mu\right)}
  \right).
  \label{eqn:gaussian:kernel:function}
\end{align}
A Gaussian approximation for $\hat{\varphi}_i(\theta)$ is
\begin{equation}
  \hat{\varphi}^\mathrm{g}_i(\theta) = K(\theta; \bar{\theta}^*_i, Q_i),
\label{eqn:gaussian:approx:to:component:density}
\end{equation}
where
\begin{align*}
  \bar{\theta}^*_i &= \frac{1}{m} \sum_{j=1}^m \theta^*_{i(j)}, \\ 
  Q_i &= \frac{1}{m-1} \sum_{j=1}^m (\theta^*_{i(j)} - \bar{\theta}^*_i) (\theta^*_{i(j)} -
  \bar{\theta}^*_i)^T,
\end{align*}
are the sample mean and sample covariance of the ABC posterior sample $\theta^*_{i(1)},
\dots, \theta^*_{i(m)}$. A consequence of using 
(\ref{eqn:gaussian:approx:to:component:density}) is that the product of the
density approximations is also Gaussian (though in general unnormalised):
\begin{equation}
  \prod_{i=2}^n \hat{\varphi}^\mathrm{g}_i(\theta) = w \cdot K(\theta; a, B), 
\label{eqn:product:of:gaussians}
\end{equation}
where
\begin{align}
   & B = \left( \sum_{i=2}^n Q_i^{-1} \right)^{-1}, \label{eqn:B:product} \\  
  &a = B\left( \sum_{i=2}^n Q_i^{-1} \bar{\theta}^*_i \right),\\
  & w = \det(2 \pi  B)^{1/2} \prod_{i=2}^n \det(2 \pi
  Q_i)^{-1/2} \times \nonumber \label{eqn:w:product} \\
  &\quad
  \prod_{s=2}^n 
  \prod_{t>s}^n
  \exp\left(
  -\tfrac{1}{2}{ (\bar{\theta}^*_s - \bar{\theta}^*_t)^T R_{st}
  (\bar{\theta}^*_s - \bar{\theta}^*_t) }
  \right), \\
  & R_{st} = Q_s^{-1} B Q_t^{-1}.
\end{align}
We note the following properties of approximation
\eqref{eqn:gaussian:approx:to:component:density} \citep[see, for example,][]{Mardia_etal79}.  If the densities
$\tilde{\varphi}_i(\theta)$ from which the  $\theta^*_{i(1)}, \dots, \theta^*_{i(m)}$ are
drawn are Gaussian, i.e., $\tilde{\varphi}_i(\theta) = K(\theta; \mu_i, \Sigma_i)$, then
$\bar{\theta}^*_i$ and $Q_i$ are unbiased and consistent estimators of $\mu_i$ and $\Sigma_i$,
respectively, and hence $a$ and $B$ are consistent estimators of the true mean and
covariance of $\prod \tilde{\varphi}_i(\theta)$.  More generally, for  
$\tilde{\varphi}_i(\theta)$ which is not necessarily Gaussian, $\bar{\theta}^*_i$ and $Q_i$ are
consistent estimators of the mean and the variance of the Gaussian density,
$\hat{\varphi}^\mathrm{g}_i(\theta)$, which minimises the Kullback--Leibler divergence,
\[
\text{KL}(\tilde{\varphi}_i(\theta)\|\hat{\varphi}^\mathrm{g}_i(\theta)) = 
\int \tilde{\varphi}_i(\theta) \log\left( \tilde{\varphi}_i(\theta) \big\slash 
\hat{\varphi}^\mathrm{g}_i(\theta)\right) {\rm d} \theta;
\]
i.e., for each $i$,
$\hat{\varphi}^\mathrm{g}_i(\theta)$ is asymptotically the ``optimal'' Gaussian
approximation to $\tilde{\varphi}_i(\theta)$.  No such relevant optimality holds for the
product of densities, however: the (normalised) product of Gaussians, each of which is
closest in the KL sense to $\tilde{\varphi}_i(\theta)$, is in general not the Gaussian
closest to (the normalised version of) $\prod\tilde{\varphi}_i(\theta)$; and indeed it may
be very substantially different.  In other words, as $m \rightarrow \infty$, $a$ and
$B$ do \emph{not} in general minimise
 \[
 \text{KL}\left( \left\{ \prod \tilde{\varphi}_i(\theta) \Big/ \int \left(\prod
 \tilde{\varphi}_i(\theta) 
 \right) \right\} \Big\| K(\theta,a,B)\right).
 \]

%We should note that if the prior $pi(\theta)$ is also Gaussian then it is trivial to
%obtain a Gaussian approximation to full posterior distribution $\pi(\theta|x)$.
%Even if on the other hand we choose non-Gaussian priors, then the normalizing constant in
%(\ref{eqn:approx:posterior}) can be calculated easily numerically.

\subsection{Kernel density estimate for $\hat{\varphi}_i(\theta)$}
\label{sec:kern-estim-hatv}

A second method we consider is to estimate each density $\tilde{\varphi}_i(\theta)$ using a
kernel density estimate (see for instance \citet{silverman:density_estimation:86} and
\citet{Wand+Jones:kernel_smoothing:95}). A kernel density
estimate based on Gaussian kernel functions \eqref{eqn:gaussian:kernel:function} is 
\begin{gather}
  \hat{\varphi}_i^\mathrm{k}(\theta) = \frac{1}{m} \sum_{j=1}^m K(\theta; \theta^*_{i(j)}, H_i ),
  \label{eqn:KDE}
\end{gather}
where $H_i$ is a bandwidth matrix.  We follow the approach of 
\cite{fukunaga} in choosing the bandwidth matrix such that the shape of the kernel mimics
the shape of the sample, in particular by taking $H_i$ to be proportional to the
sample covariance matrix, $Q_i$.  Using bandwidth matrix
\begin{equation}
  H_i = q \cdot m^{-2/(d+4)} \, Q_i,
  % H_i = \left\{ m \left( d+2 \right)/4 \right\}^{-2/(d+4)} \, Q_i,
\label{eqn:bandwidth}
\end{equation}
where $q>0$ is a constant not dependent on $m$, ensures desirable behaviour as the sample size
$m \rightarrow \infty$.  In particular, in terms of the little-o notation ($a_m = o(b_m)$
as $m\rightarrow \infty$ denotes $\lim_{m\rightarrow \infty} | a_m/b_m |=0$) and with $E$
denoting expectation, using choice of bandwidth (\ref{eqn:bandwidth}), subject to
mild regularity conditions on $\tilde{\varphi}_i(\theta)$ 
\citep{Wand+Jones:kernel_smoothing:95},
\begin{align}
  & E \left\{ \hat{\varphi}^\mathrm{k}_i(\theta) \right\} = \tilde{\varphi}_i(\theta) + o(1),
  \label{eqn:asymp1}\\
  & E \left\{ \hat{\varphi}^\mathrm{k}_i(\theta)^2 \right\} = \tilde{\varphi}_i(\theta)^2 + o(1).
  \label{eqn:asymp2}
\end{align}
From (\ref{eqn:asymp1})--(\ref{eqn:asymp2}),
the bias, b$\{\hat{\varphi}^\mathrm{k}_i(\theta)\} = E \left\{
\hat{\varphi}^\mathrm{k}_i(\theta) \right\} -
\tilde{\varphi}_i(\theta)$, the variance, var$\{\hat{\varphi}^\mathrm{k}_i(\theta)\} = E \left\{
\hat{\varphi}^\mathrm{k}_i(\theta)^2 \right\}  - E
\left\{\hat{\varphi}^\mathrm{k}_i(\theta)\right\}^2$, and
the mean integrated squared error,
\begin{align}
  \text{MISE}\{ \hat{\varphi}^\mathrm{k}_i \} & = E \int
  \left(\hat{\varphi}^\mathrm{k}_i(\theta) - 
  \tilde{\varphi}_i(\theta) \right)^2 \mathrm{d} \theta, % \\
%  & = \int \left( \text{var}\right\{\hat{\varphi}_i(\theta)\left\} +
%  \text{b}\{\hat{\varphi}_i(\theta)\}^2 \right) \mathrm{d} \theta
\label{eqn:MISE}
\end{align}
are all $o(1)$.  These results generalise routinely to the case of a product of $n$ kernel density
estimates, that is, in which $\prod \hat{\varphi}^\mathrm{k}_i(\theta)$ is used 
as an estimator for $\prod
\tilde{\varphi}_i(\theta)$.  It follows that since the $\theta^*_{i(j)}$ are independent for all
$i,j$, then, using (\ref{eqn:asymp1})--(\ref{eqn:asymp2}), 
\begin{align*}
  &\text{b}\left\{\prod \hat{\varphi}^\mathrm{k}_i(\theta)\right\} = \left\{ \prod E \,
  \hat{\varphi}^\mathrm{k}_i(\theta) \right\} - \prod \tilde{\varphi}_i(\theta) = o(1), \\
  & \text{var}\left\{\prod \hat{\varphi}^\mathrm{k}_i(\theta)\right\} = \prod E \left\{
  \hat{\varphi}^\mathrm{k}_i(\theta)^2 \right\} - \prod \left\{ E 
  \hat{\varphi}^\mathrm{k}_i(\theta) \right\} ^2 = o(1), \\ 
  & \text{MISE}\left\{ \prod \hat{\varphi}^\mathrm{k}_i \right\} = E \int \left( \prod
  \hat{\varphi}^\mathrm{k}_i(\theta) -
\prod \tilde{\varphi}_i(\theta) \right)^2 \mathrm{d} \theta = o(1).
\end{align*}
Hence, in the sense defined by the latter equation, the density estimator 
$\prod \hat{\varphi}^\mathrm{k}_i(\theta)$ converges to the true density $\prod
\tilde{\varphi}_i(\theta)$ as
$m \rightarrow \infty$.
% (\ref{eqn:approx:posterior}) with kernel density estimates (\ref{eqn:KDE})
% and bandwidth (\ref{eqn:bandwidth}), converges to the true posterior density
% $\pi(\theta\vert x)$ as $m \rightarrow \infty$.

Regarding the choice of $q$ in (\ref{eqn:bandwidth}), in certain settings it is possible
to determine an optimal value.  Suppose that the true density $\tilde{\varphi}_i(\theta)$
is Gaussian and let $\hat{\varphi}^\mathrm{k}_i(\theta)$ in (\ref{eqn:KDE}) be a kernel density
estimate of $\tilde{\varphi}_i(\theta)$.  Then 
\begin{equation}
q = \{\left( d+2 \right)/4\}^{-2/(d+4)}
\label{eqn:optimal:q}
\end{equation}
is optimal in the sense that (\ref{eqn:bandwidth}) is then an unbiased and consistent
estimator of the bandwidth that minimises the leading term of the large-$m$ asymptotic
expansion of (\ref{eqn:MISE}); see \citet[][p111]{Wand+Jones:kernel_smoothing:95}.
Analogous calculations are rather more involved in the product case, however: even with
the assumption that each $\tilde{\varphi}_i(\theta)$ is Gaussian, no closed expression for
$q$ is possible.  Hence, in the examples in the following section, \S\ref{sec:examples}, we
opted to tune $q$ in the heuristic way described by \citet{Wand+Jones:kernel_smoothing:95},
starting with a large $q$ (ten times that in \eqref{eqn:optimal:q}) then reducing it
manually until
``random'' fluctuations begin to appear in the density estimates.

A consequence of using Gaussian kernel functions \eqref{eqn:gaussian:kernel:function} in
(\ref{eqn:KDE}) is that the product of the density approximations is then itself a
weighted mixture of $(n-1)^m$ Gaussians,
\begin{align}
  & \prod_{i=2}^n \hat{\varphi}^\mathrm{k}_i(\theta)  
= m^{(1-n)}\prod_{i=2}^n \sum_{j=1}^m K(\theta; \theta^*_{i(j)}, H_i ) \nonumber \\
& = m^{(1-n)} \sum_{j_2, \ldots, j_n}^m \prod_{i=2}^n K(\theta; \theta^*_{i(j_i)}, H_i ) \nonumber \\
& = \sum_{j_2,\ldots,j_n}^m w_{j_2,\ldots,j_n} K(\theta; a_{j_2,\ldots,j_n},
B_{j_2,\ldots,j_n}), \label{eqn:prod:phi:kernel}
\end{align}
where expressions for the covariances $B_{j_2,\ldots,j_n}$, means $a_{j_2,\ldots,j_n}$,
and weights $w_{j_2,\ldots,j_n}$, analogous to those in
(\ref{eqn:B:product})--(\ref{eqn:w:product}), are given in Appendix 1.

\subsection{Estimating the posterior density}

Sections \S\ref{sec:gauss-appr-hatv} and \S\ref{sec:kern-estim-hatv} describe methods for
computing the factor $\prod \hat{\varphi}_i(\theta)$ in \eqref{eqn:approx:posterior}.  For
calculating an estimate of the full posterior, $\hat{\pi}(\theta \vert x)$ in
\eqref{eqn:approx:posterior}, we must multiply $\prod \hat{\varphi}_i(\theta)$ by
$\pi(\theta)^{(2-n)}$ and normalise.  Let us suppose that the prior is Gaussian,
$\pi(\theta) = K(\theta; \mu_{\text{pri}}, \Sigma_{\text{pri}})$.  For the case where we
are using the Gaussian approximation, $\hat{\varphi}^\mathrm{g}_i(\theta)$ from
\eqref{eqn:gaussian:approx:to:component:density},  for each $\hat{\varphi}_i(\theta)$,
then the posterior is
\begin{equation}
  \hat{\pi}^\mathrm{g}(\theta \vert x) = K(\theta; \mu_{\text{post}},
   \Sigma_{\text{post}}),
   \label{eqn:post:gaussian}
\end{equation}
where
\begin{align}
   \Sigma_{\text{post}} & = \left( (2-n) \Sigma_\text{pri}^{-1} + B^{-1} \right)^{-1},
   \label{eqn:post:Sigma}\\
  \mu_{\text{post}} & = \Sigma_{\text{post}} \left( (2-n) \Sigma_\text{pri}^{-1} \, \mu_{\text{pri}}
  + B^{-1} a\right), \label{eqn:post:mu}
\end{align}
and $a$ and $B$ are as defined in \eqref{eqn:product:of:gaussians}. 

If instead we use the kernel approximation, $\hat{\varphi}^\mathrm{k}_i(\theta)$ from
\eqref{eqn:KDE}, for each $\hat{\varphi}_i(\theta)$, then the posterior density is 
\begin{align}
  \hat{\pi}^\mathrm{k}(\theta \vert x) = \sum_{j_2,\ldots,j_n}^m w'_{j_2,\ldots,j_n}
& K(\theta; a'_{j_2,\ldots,j_n},
B'_{j_2,\ldots,j_n}) \Big\slash \nonumber \\ 
& \sum_{j_2,\ldots,j_n}^m w'_{j_2,\ldots,j_n},
\label{eqn:post:kernel} \end{align}
where expressions for $B'_{j_2,\ldots,j_n}$, $a'_{j_2,\ldots,j_n}$ and $w'_{j_2,\ldots,j_n}$
are in the Appendix.

\subsection{An expression for the posterior density}
\label{sec:PWABC:posterior}

In the preceding sections we considered how to sample from the $\varphi_i(\theta)$ and
then use the samples to estimate the posterior density $\pi(\theta|x)$.  Here we
consider in more detail the implied posterior density which is targeted by PW-ABC.  
%To keep presentation simple and to enable clear comparison between PW-ABC and ABC, we
%take $s(\cdot) = \text{Identity}(\cdot)$, and assume that the $x_i$ are realisations of
%a continuous, scalar random variable $X(t)$, but discuss at the end of this section how
%the interpretation is changed little if the $x_i$ are discrete and/or non-scalar.
For either of PW-ABC and ABC, the posterior can be written as
\begin{equation}
   \tilde{\pi}(\theta | x) \propto \tilde{\pi}(x \vert \theta) \pi(\theta),
   \label{eqn:ABC:posterior}
\end{equation}
where $\tilde{\pi}(x \vert \theta)$ is, respectively, either the implied PW-ABC or ABC
approximation to the likelihood.  First, we define the function
 \begin{equation}
    K_{\varepsilon,p}(z) = V^{-1} 
    \mathds{1} \{\| z \|_p \leq \varepsilon \}, 
   \label{eqn:ABC:lik:approx:kernel}
 \end{equation}
where argument $z$ is of dimension, say, $u$, and either continuous- or
discrete-valued in accord with the support of the data; $\| \cdot \|_p$ is the $L^p$-norm;
$\mathds{1}\!\left\{\cdot \right\}$ is an indicator function; and $V$, which depends on
$u$, $\varepsilon$, and $p$, is such that $\int K_{\varepsilon,p}(z) {\mathrm d} z = 1$,
with this integral interpreted as a sum in the discrete case.  For ABC with distance
$d(\cdot,\cdot)$ taken to be the $L^p$-norm of the difference between its arguments, the
implied ABC approximation to the likelihood \citep{Wil13} is the convolution
 \begin{align}
    &\tilde{\pi}_{\text{ABC}}(x \vert \theta) = \int \pi(y \vert \theta) 
    K_{\varepsilon,p} \left( y - x \right) {\mathrm d} y. 
 %  & = \int \pi(y_2 \vert x_{1}, \theta) 
 %  \left[ \prod_{i=3}^n \pi(y_i \vert y_{i-1}, \theta) \right] 
 %  K_{\varepsilon,2} \left( y - x \right) \rm d y.
   \label{eqn:ABC:lik:approx}
 \end{align}
 Hence ABC replaces the true likelihood with an approximate version averaged over
 an $L^p$-ball of radius $\varepsilon$ centred on the data vector, $x$.
 In PW-ABC, we target each $\varphi_i(\theta)$ by an ABC approximation 
 $\tilde{\varphi}_i(\theta) \propto \tilde{\pi}_{\text{ABC}}(x_i
 \vert x_{i-1}, \theta) \pi(\theta)$, with
 \begin{equation*}
    \tilde{\pi}_{\text{ABC}}(x_i \vert x_{i-1}, \theta)  =
    \int \pi(y_i \vert x_{i-1} , \theta)  K_{\varepsilon,p} \left( y_i - x_i \right)
    {\mathrm d} y_i,
  %  \label{eqn:ABC:factor:lik:approx}
 \end{equation*}
 and the implied PW-ABC likelihood is the product
 \begin{equation}
   \tilde{\pi}_{\text{PW-ABC}}(x \vert \theta) = \prod \tilde{\pi}_{\text{ABC}}(x_i \vert
   x_{i-1}, \theta).
   \label{eqn:PWABC:lik:approx}
 \end{equation}
% and hence we can write the implied PW-ABC approximation
 %Since we are assuming that $x_i$ is scalar and continuous, $K_{\varepsilon,p} \left( y_i - x_i
 %\right)$ reduces to $ (2 \varepsilon)^{-1} \mathds{1}\{|y_i - x_i| \leq \varepsilon\}.$
 
 Now, to compare directly the implied ABC and PW-ABC likelihood approximations, we
 neglect as before the likelihood contribution from the first observation $x_1$, then
 denote by $x'$ the vector $x$ with $x_1$ removed (and similar for $y$); hence
 we can write (\ref{eqn:ABC:lik:approx}) and (\ref{eqn:PWABC:lik:approx}), respectively,
 as
%\begin{align} & \tilde{\pi}_{\text{PW-ABC}}(x \vert \theta) = \prod_{i=2}^n
%\tilde{\pi}_{\text{ABC}}(x_i \vert x_{i-1}, \theta) \nonumber \\ & = \int \left[
%\prod_{i=2}^n \pi(y_i \vert x_{i-1}, \theta) \right] K_{\varepsilon,1} \left( y - x
%\right) \rm d y.  \end{align}
\begin{equation}
   \int \pi(y_2 \vert x_{1}, \theta) 
  \left[ \prod_{i=3}^n \pi(y_i \vert y_{i-1}, \theta) \right] 
  K_{\varepsilon,p} \left( y' - x' \right) {\mathrm d} y',
\end{equation}
and
\begin{equation}
   \int \pi(y_2 \vert x_{1}, \theta) 
  \left[ \prod_{i=3}^n \pi(y_i \vert x_{i-1}, \theta) \right] 
  K^*_{\varepsilon,p} \left( y' - x' \right) {\mathrm d} y',
  \label{eqn:PWABC:lik:approx2}
\end{equation}
where
\begin{equation}
  K^*_{\varepsilon,p} \left( z' \right) = \prod_{i=2}^n K_{\varepsilon,p} \left( z'_i
  \right).
  \label{eqn:PWABC:implied:convolution:kernel}
\end{equation}
Two differences between ABC and PW-ABC are clear: first, in ABC the conditioning is on the
simulated trajectory, whereas in PW-ABC the conditioning is on the data; and second, in
PW-ABC the convolution is with respect to a different kernel
(\ref{eqn:PWABC:implied:convolution:kernel}).  This implied kernel seems intuitively
reasonable; for example, if the $x_i$ are scalar then the convolution in
(\ref{eqn:PWABC:lik:approx2}) amounts to an averaging over a hypercube of side length $2
\varepsilon$ centred on $x'$.  The difference in the shapes of the regions defined by
$K_{\varepsilon,p}(\cdot)$ and $K^*_{\varepsilon,p}(\cdot)$ is of secondary importance,
however, since PW-ABC enables use of a much smaller $\varepsilon$ than ABC, so the
averaging will be over a much smaller region around $x'$, and the approximate
likelihood will typically be much closer to the true.

%not in general the same, although they do coincide
%if $p=1$ (i.e., the $L^1$-norm) is used in the ABC in (\ref{eqn:ABC:lik:approx}).  

% If the $x_i$ are discrete rather than continuous then the foregoing comparison is
% not substantially changed, except that the kernel (\ref{eqn:ABC:lik:approx:kernel}) is taken to have
% support in accordance with the $x_i$, and the normalising constant is instead 
% $V = \sum_z K_{\varepsilon,p}(z)$, where the sum is over the support.
% 
% If the $x_i$ are non-scalar then the kernel for PW-ABC in (\ref{eqn:PWABC:lik:approx}) is
% $K_{\varepsilon,1}(\cdot)$ only if $p=1$ is used in (\ref{eqn:ABC:factor:lik:approx}) for
% each $\tilde{\varphi_i}(\theta)$.  Using $p \neq 1$ leads to neither
% $K_{\varepsilon,p}(\cdot)$ nor $K_{\varepsilon,1}(\cdot)$ in (\ref{eqn:PWABC:lik:approx}) 
% but instead a hybrid (though intuitively reasonable) kernel.
% 
% The different kernels define different shaped regions over which the convolution averages.
% More important than differences in shape, however, is the key point that PW-ABC will
% enable use of a much smaller $\varepsilon$ than ABC, so the averaging will be over a much
% smaller region around $x$, typically making the implied approximate likelihood much closer to the
% true.

\section{Some other considerations}

\subsection{Estimating the marginal likelihood}
In some applications, especially when model comparison is of interest,
it is useful to compute the marginal likelihood of the data given the model.
The marginal likelihood is
\begin{align}
  \pi(x) & = \int \pi(x \vert \theta) \pi(\theta) \mathrm{d} \theta
  \label{eqn:marg:lik:true}
\\
  & = \left( \prod_{i=2}^n c_i \right) \int \left( \prod_{i=2}^n
  \varphi_i(\theta) 
  \right) \pi(\theta)^{2-n} \mathrm{d} \theta.
  \label{eqn:marg:lik:true:factorised}
%  & \approx \left( \prod_{i=2}^n \hat{c}_i \right) \int w_\text{post} K(\theta;
%  \mu_\text{post}, \Sigma_\text{post}) \mathrm{d} \theta,  \nonumber %\\
%  & \approx \prod_{i=2}^n \hat{c}_i \int \hat{\varphi}_i(\theta) \pi(\theta)^{2-n} \mathrm{d} \theta
\end{align}
The unknown $c_i$ can be estimated by $\hat{c}_i = m/(V\,M_i)$, where
$M_i$ equals the number of ABC draws necessary in the $i$th interval to achieve $m$
acceptances, 
and $V$ is defined in (\ref{eqn:ABC:lik:approx:kernel}); see
Appendix 2. % Replacing the
% $\varphi_i(\theta)$ with the Gaussian approximations $\hat{\varphi}_i(\theta)$ and 
For the integral in \eqref{eqn:marg:lik:true:factorised}, using the Gaussian approximation
\eqref{eqn:product:of:gaussians} leads to
\begin{align} 
&  \int \left( \prod_{i=2}^n
\hat{\varphi}^\mathrm{g}_i(\theta) \right) \pi(\theta)^{2-n} \mathrm{d} \theta \nonumber \\
 &  =  w \cdot (\text{det} B)^{-1/2} \cdot (\text{det} \Sigma_{\text{post}})^{1/2}  
 \cdot ( \text{det} (2 \pi \Sigma_\text{pri}))^{(n/2 - 1)} \times \nonumber \\
 & \exp{\left\{ -\frac{1}{2}(a-\mu_\text{pri})^T \left( 
 (2-n)^{-1}\Sigma_\text{pri} + B \right)^{-1} (a-\mu_\text{pri}) \right\}},
 \label{eqn:marg:lik:gaussian:approx}
 \end{align}
 whereas using the kernel approximation \eqref{eqn:KDE} gives
 \begin{equation}
\int \left( \prod_{i=2}^n
\hat{\varphi}^\mathrm{k}_i(\theta) \right) \pi(\theta)^{2-n} \mathrm{d} \theta =  
\sum_{j_2,\ldots,j_n}^m w'_{j_2,\ldots,j_n}.
\label{eqn:marg:lik:kernel:approx}
\end{equation}

\subsection{Practical numerical calculations for the kernel approximation}
\label{sec:pract-numer-calc}

Since expressions \eqref{eqn:prod:phi:kernel}, \eqref{eqn:post:kernel},
\eqref{eqn:marg:lik:kernel:approx} for the kernel case involve sums with $(n-1)^m$ terms,
these expressions are largely of academic interest and are typically not suitable for
practical calculations.  For the examples in this paper we used a more direct numerical
approach, first writing \eqref{eqn:g:in:expression:for:posterior} as
\[
g(\theta) = \exp \left( \sum_{i=2}^n h_i(\theta) \right) \pi(\theta),
\]
where $h_i(\theta) = \log \left( \varphi_i^\mathrm{k}(\theta) \slash \pi(\theta) \right)$,
and then evaluating $h_i(\theta)$, $\pi(\theta)$ and hence $g(\theta)$ pointwise on a fine
lattice.  Performing calculations in this way on the log scale avoids underflow errors and
improves numerical stability compared with trying to evaluate
\eqref{eqn:g:in:expression:for:posterior} directly.  As a further check for robustness, we
varied the lattice position and resolution to make sure the results were insensitive to
the particular choices.

\subsection{Sampling from the posterior distribution}
\label{sec:sampling:from:posterior}
In some circumstances it may be desirable to draw samples
from the approximate posterior density.  In the Gaussian case, drawing from
\eqref{eqn:post:gaussian} is straightforward.  For the kernel case,
\eqref{eqn:post:kernel}, in principle sampling
can be achieved by normalising the weights, randomly 
choosing a component with probability equal to these normalised weights, then sampling from the
selected Gaussian component.  But in practise, again, the large number of
terms in \eqref{eqn:post:kernel} can preclude this approach.  Other possibilities
include using a Gibbs sampler, or sampling approximately using Gaussian mixtures with
fewer components; see \cite{ihler}.

\section{Examples} \label{sec:examples}

In this section we test PW-ABC on synthetic data from four models.  The first, as a toy
illustrative example, involves inferring from IID data the probability of success in a
binomial experiment.  Second is the Cox-Ingersoll-Ross model, a 
stochastic-differential-equation model for which the continuous state variable has known
transition density, which we use to investigate PW-ABC with $\varepsilon>0$.  
Third, we consider an integer-valued time series model called
INAR(1), a model for which the likelihood is available (albeit awkward to compute) and
enables comparison of our approach with the ``gold standard'' MCMC approach.  Finally, we
consider a stochastic Lotka--Volterra model, a simple example from a common class of
models (which occur, for instance, in modelling stochastic chemical kinetics) in which the
likelihood, and therefore many standard methods of inference, are unavailable.  The
datasets for each example are given in the Supplementary Material.

\subsection{Binomial model} \label{sec:binomial:model}
\begin{figure}[!t]
 \centering
 \includegraphics[angle=0, width=8cm]{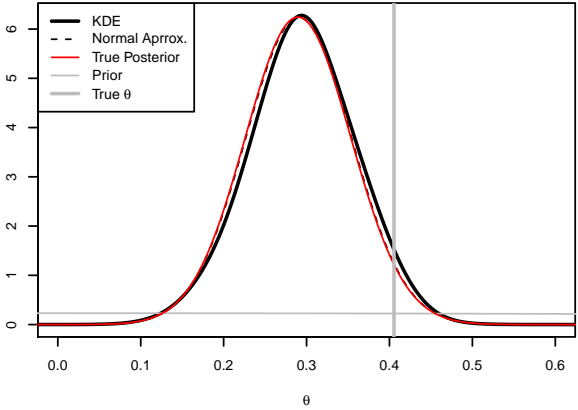} 
 \caption{Results for the binomial model in  \S\ref{sec:binomial:model}.  Shown are 
 the true posterior density, ${\pi}(\theta|x)$, the posterior
 density approximations $\hat{\pi}^\mathrm{g}(\theta|x)$ and 
 $\hat{\pi}^\mathrm{k}(\theta|x)$, the prior, and the true $\theta$.}
 \label{fig:binomial}
\end{figure}

For this toy example we suppose the data is the set $x = \{x_1, \ldots, x_{10}
\}$ of $n=10$ observations from the model $X_i\sim\text{Binom}(k_i=100,p=0.6)$.  We
work in terms of the transformed parameter $\theta = \textrm{logit} (p)$, 
using a prior $\pi(\theta) \sim N(0,3^2)$.  For this model the data are IID,
so that $\pi(x_i \vert x_{i-1}, \theta) = \pi(x_i \vert \theta)$.  Exact samples from
$\varphi_i(\theta)$ can be obtained by sampling $\theta^*$ from the prior, sampling
$X_i^*\sim\text{Binom}(100,\theta^*)$, and then accepting $\theta^*$ if and only if $X_i^*
= x_i$. 
%(For this toy, of course, alternatively samples could be drawn directly from the known
%posterior.) For this toy example we suppose the data is the set $x = \{x_1,
%\ldots, x_{10} \}$ of $n=10$ observations from the model
%$X_i\sim\text{Binom}(k_i=100,p=0.6)$.  We use the conjugate prior $\pi(p) \sim
%\text{Beta}(\alpha=6,\beta=10)$ so the true posterior is $\pi(p|x) \sim
%\text{Beta}(\alpha=6+\sum{}x_i,\beta=10+\sum{}k_i-x_i)$.  For this model the data are
%IID, so that $\pi(x_i \vert x_{i-1}, p) = \pi(x_i \vert p)$.  In addition, it is trivial
%to show that and $\varphi_i(p) \propto p^{\alpha + x} (1-p)^{10 - x + \beta}$, i.e.
%$p|x_i \sim \mbox{Beta}(\alpha + x_i,10 - x_i + \beta)$ and hence, exact samples from
%$\varphi_i(p)$ can be drawn in a straightforward manner without necessarilty resorting to
%ABC. 
We follow the PW-ABC approach described in Section~\ref{sec:methods}, drawing $m=5000$
samples from each $\varphi_i(\theta)$, using these samples to construct Gaussian
$\hat{\varphi}^\mathrm{g}_i(\theta)$ and kernel density
$\hat{\varphi}^\mathrm{k}_i(\theta)$ approximations, then using these density
approximations to construct approximate posterior densities,
$\hat{\pi}^\mathrm{g}(\theta|x)$ and $\hat{\pi}^\mathrm{k}(\theta|x)$.
Figure \ref{fig:binomial} shows that the approximate posterior densities are very close to
the true posterior density for this example.  The true log marginal likelihood,
$\log \pi(x)$, computed 
by direct numerical integration of \eqref{eqn:marg:lik:true}, is $-31.39$; 
using approximation $\hat{\varphi}^\mathrm{g}_i(\theta)$ and
\eqref{eqn:marg:lik:gaussian:approx} gives $-31.44$; and using approximation
$\hat{\varphi}^\mathrm{k}_i(\theta)$ and numerical integration of the left-hand side of
\eqref{eqn:marg:lik:kernel:approx} gives $-31.48$.

\subsection{Cox--Ingersoll--Ross Model}
\label{sec:CIR}

The Cox-Ingersoll-Ross (CIR) model \citep{MR785475} is a stochastic differential equation
describing evolution of an interest rate, $X(t)$.  The model is
\begin{equation*}
  \mathrm{d}X(t) = a(b - X(t))\mathrm{d}t + \sigma\sqrt{X(t)}\mathrm{d}W(t),
\end{equation*}
where $a$, $b$ and $\sigma$ respectively determine the reversion speed, long-run value
and volatility, and where $W(t)$ denotes a standard Brownian motion.  The density of
$X(t_i)|X(t_j),a,b,\sigma$ ($t_i>t_j$) 
is a non-central chi-square \citep[eq.~18]{MR785475}, and hence
the likelihood is known in closed form.  In such a situation (PW-)ABC is unnecessary; however,
we include the CIR model here 
as a simple example of PW-ABC applied to a problem with a continuous
state variable, where non-zero choice of $\varepsilon$ is necessary, and where 
the true posterior distribution is available for comparison.

We generated $n=10$ equally spaced observations from a CIR process with parameters
$(a,b,\sigma)=(0.5,1,0.15)$ and $X(0)=1$ on the interval $t \in [0,4.5]$. Treating $a$
and $\sigma$ as known, we performed inference on the transformed parameter
$\theta=\log(b)$ with a Uniform prior on the interval $(-5, 2)$.  Using
$\varepsilon=10^{-2}$ we drew samples of size $m=10,000$ for each $\varphi_i(\theta)$,
$i=1,\ldots,9$, achieving acceptance rates around 1.5\% on average.

\begin{figure} 
   \centering 
   \includegraphics{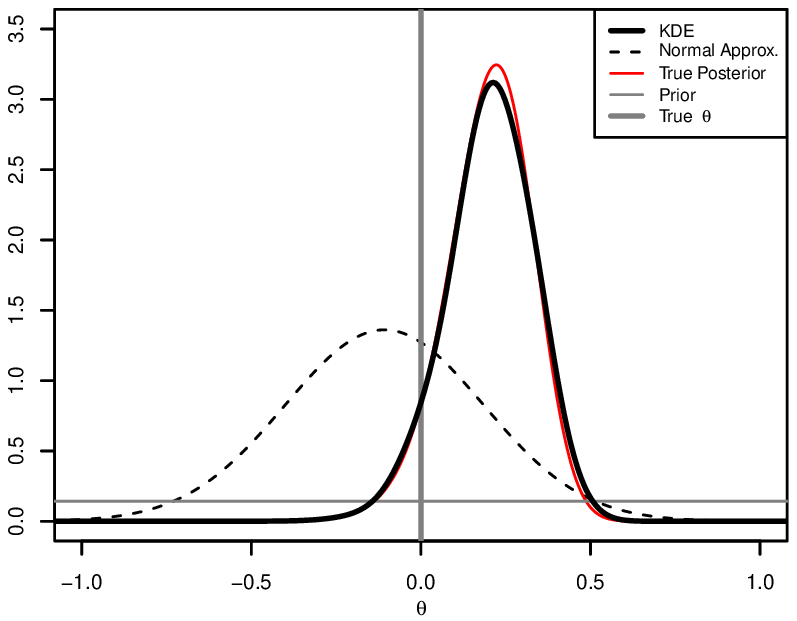}
   \includegraphics{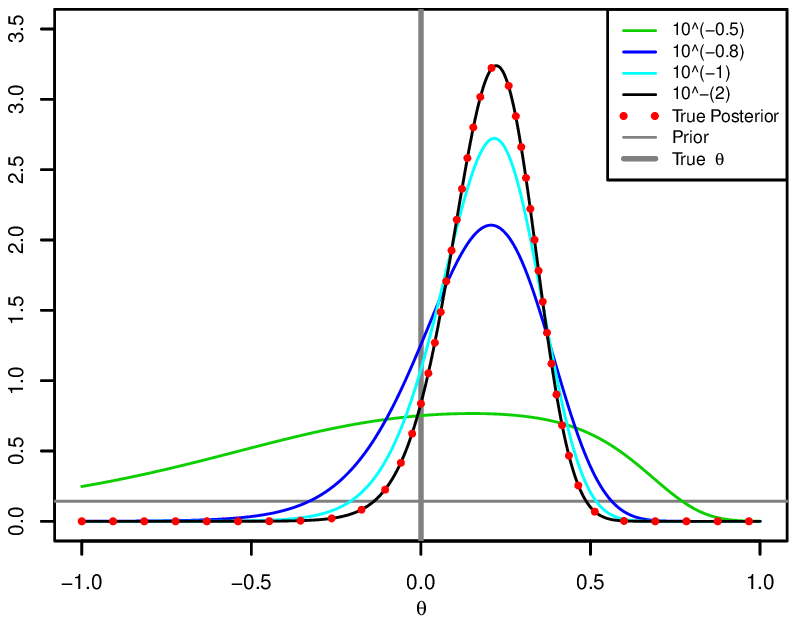} 
   \caption{Results for the CIR model of \S\ref{sec:CIR}.  The
   top plot shows the true posterior density, ${\pi}(\theta|x)$; the PW-ABC
   posterior density approximations $\hat{\pi}^\mathrm{g}(\theta|x)$ and
   $\hat{\pi}^\mathrm{k}(\theta|x)$ using $\varepsilon = 10^{-2}$; the prior; and the
   true $\theta$.  The bottom plot shows, for various
   values of $\varepsilon$, the true PW-ABC posterior (as defined in 
   \S\ref{sec:PWABC:posterior}).}
   \label{fig:CIR}
\end{figure}

Figure \ref{fig:CIR} 
shows the true posterior density, $\pi(\theta|\mathcal{X})$, together with the
Gaussian- and kernel-based PW-ABC approximations, $\hat{\pi}^\mathrm{g}(\theta|x)$ and
$\hat{\pi}^\mathrm{k}(\theta|x)$. The Figure
shows that the kernel approximation $\hat{\pi}^\mathrm{k}(\theta|x)$ agrees very
well with the true posterior.  The Gaussian approximation
$\hat{\pi}^\mathrm{g}(\theta|x)$ does badly here, which is due to skewness of  
the densities $\varphi_i(\theta)$.

For this example, estimates of the log marginal
likelihood, $\log \pi(\mathcal{X})$ are as follows: by direct numerical
integration of \eqref{eqn:marg:lik:true}, $8.14$; using approximation
$\hat{\varphi}^\mathrm{g}_i(\theta)$, $2.78$; and by using
$\hat{\varphi}^\mathrm{k}_i(\theta)$  in conjunction with numerical integration of the
left-hand side of \eqref{eqn:marg:lik:kernel:approx}, $7.93$.

\subsection{An integer-valued autoregressive model}
\label{sec:appl-INAR}

Integer-valued time series arise in contexts such as modelling monthly traffic fatalities
\citep{NealRao07} or the number of patients in a hospital at a sequence of time points
\citep{Morina+et_al:model_hospital_admissions:11}. Consider the following integer-valued
autoregressive model of order $p$, known as INAR$(p)$:
\begin{equation}
  X_t = \sum_{i=1}^p \alpha_i \circ X_{t-i} + Z_t, \quad t \in \mathbb{Z},
\end{equation}
where $Z_{t}$ for $t>1$ are independent and identically distributed integer-valued random
variables with $E[Z_t^2] < \infty$, with the $Z_t$ assumed to be independent of the $X_t$.
Here we assume $Z_t \sim Po(\lambda)$.  Each operator $\alpha_i \circ$ denotes binomial
thinning defined by
\begin{equation}
  \label{eq:1}
  \alpha_i \circ W = \left\{
    \begin{array}{cc}
      \mbox{Binomial}(W,\alpha_i), \quad & W > 0, \\
      0, & W = 0,
    \end{array}
  \right. 
\end{equation}
for non-negative integer-valued random variable $W$. The operators $\alpha_i\circ$,
$i=1,\ldots p$, are assumed to be independent.

\begin{figure}[!t] \centering \includegraphics[angle=270,width=7.2cm]{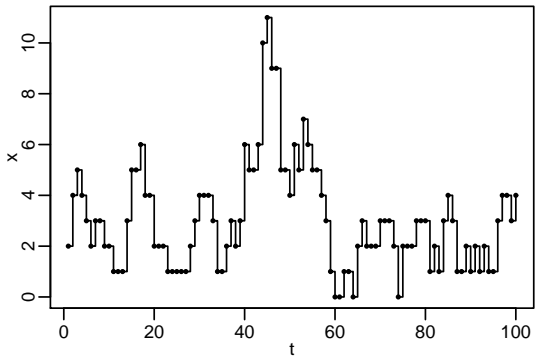}
  %-90 for eps 
  \caption{The realisation of an INAR(1) process used in the example of
  \S\ref{sec:appl-INAR}, of length $n=100$, generated using $\alpha=0.7$ and
  $\lambda=1.0$.}
 \label{fig:INAR_obs}
\end{figure}

We consider the simplest example of this model, INAR(1) \cite[see, for
example,][]{Al-OsAl87}, supposing that we have some observed data \linebreak $x
= \{x_1, \ldots, x_n\}$ from this model and wish to make inference for the parameters
$(\alpha,\lambda)$.
%The likelihood function is
%\begin{multline*}
%\pi(x | \theta) = \\
%Pr(X_1 = x_1) \cdot \prod_{t=2}^{n}Pr(X_t = x_t|X_{t-1}=x_{t-1}),
%\end{multline*}
%in which
%\begin{multline*}
% Pr(X_1 = x_1) = \frac{(\lambda/(1-\alpha))^{x_0}}{x_0!}\exp{(-\lambda/(1-\alpha))}.
%\end{multline*}
%and
%\begin{multline*}
% Pr(X_t = x_t | X_{t-1}=x_{t-1}) = \\
%\sum_{i=0}^{\min(x_t,x_{t-1})} \frac{\lambda^{x-i}}{(x-i)!}{x_{t-1} \choose i} \alpha^i
%(1-\alpha)^{x_{t-1} - i};
%\end{multline*}
%this latter expression coming from the convolution of a Poisson with a
%binomial random variable \cite[see][]{ShumGur60}.
We generated $n=100$ observations from an INAR(1) process using parameters
$(\alpha,\lambda) = (0.7, 1)$ and $X(0)=10$; the realisation is plotted in
Figure~\ref{fig:INAR_obs}. Working in terms of the transformed parameter, ${\theta} =
(\theta_1,\theta_2) = 
(\textrm{logit}(\alpha),\log (\lambda))$, we used a prior of $\mathrm{Norm}(0,3^2)$ for 
each of $\theta_1$ and $\theta_2$. For the EBC algorithm, the probability of acceptance
is around $10^{-100}$ (as estimated from PW-ABC calculations described below), 
which is prohibitively small; even the ABC algorithm requires a
value of $\varepsilon$ so large that sequential approaches (e.g., SMC-ABC) are needed.

\begin{figure}[!t] \centering \includegraphics[angle=270, width=8cm]{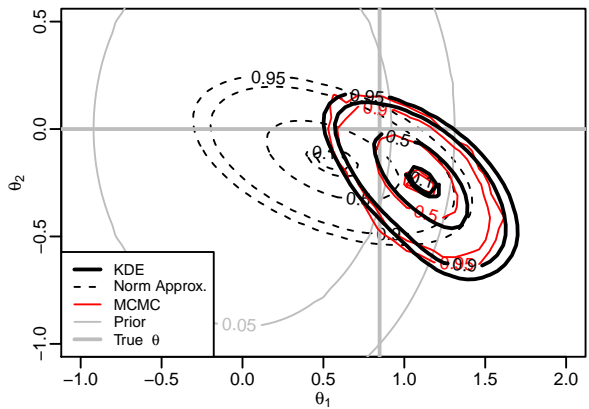}
   \caption{Results for the INAR(1) example of \S\ref{sec:appl-INAR}.  Shown
   are an MCMC approximation to the posterior density, ${\pi}(\theta|x)$, the
   posterior density approximations $\hat{\pi}^\mathrm{g}(\theta|x)$ and
   $\hat{\pi}^\mathrm{k}(\theta|x)$, the prior, and the true $\theta$.  The
   numbers on the contours denote the probability mass that they contain. }
  \label{fig:INAR_joint}
 \end{figure}

\begin{figure}[!t]
 \centering
 \includegraphics[angle=270, width=8cm]{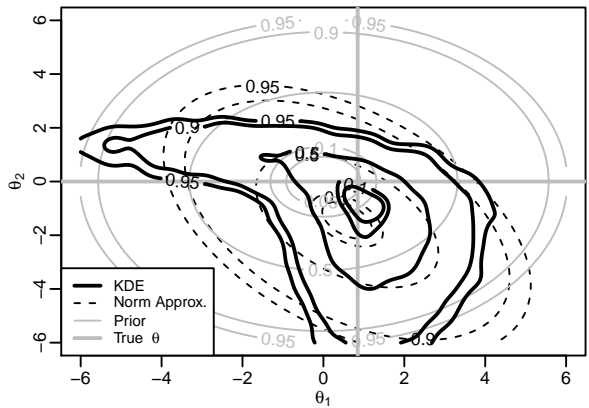} %-90 for eps
 \caption{For the INAR(1) example, an example of a factor with a ``non-Gaussian'' density:
 here $\hat{\varphi}^\textrm{g}_{50}({\theta})$ and
 $\hat{\varphi}^\textrm{k}_{50}({\theta})$ are substantially different from each other. }
 \label{fig:INAR_failure}
\end{figure}

Using PW-ABC with $\varepsilon=0$ we were able to
draw exact samples from $\varphi_i(\theta)$ for all of the $i=2,\ldots,100$ factors, and
achieve acceptance rates of around 9\%, on average.   Figure~\ref{fig:INAR_joint}
shows an estimate of the posterior density, $\pi(\theta|x)$ based on a
gold-standard MCMC approach, together with Gaussian- and kernel-based PW-ABC
approximations, $\hat{\pi}^\mathrm{g}(\theta|x)$ and
$\hat{\pi}^\mathrm{k}(\theta|x)$, with $m=10,000$ samples for each
$\varphi_i(\theta)$.  The Figure shows good agreement between the MCMC posterior and the
kernel approximation, $\hat{\pi}^\mathrm{k}({\theta}|x)$, but again somewhat 
poor agreement with the Gaussian approximation $\hat{\pi}^\mathrm{g}({\theta}|x)$.  The poor
performance of $\hat{\pi}^\mathrm{g}({\theta}|x)$ is caused by some
of the densities $\varphi_i({\theta})$ being substantially different from Gaussian; see
Figure \ref{fig:INAR_failure} which shows $\hat{\varphi}^\textrm{g}_{50}({\theta})$ and
$\hat{\varphi}^\textrm{k}_{50}({\theta})$, for example.  Using Gaussian approximations to non-Gaussian $\varphi_i({\theta})$
appears to have a strong impact on the accuracy of approximation
$\hat{\pi}^\mathrm{g}({\theta}|x)$, even, as in the present case, where the true
posterior $\pi({\theta}|x)$, and most of individual $\varphi_i({\theta})$,
are reasonably close to a Gaussian (cf. Fig.
\ref{fig:INAR_joint}).

For this example, estimates of the log marginal likelihood, $\log
\pi(x)$, are as follows: by direct numerical integration of \eqref{eqn:marg:lik:true}, $-161.1$; using approximation $\hat{\varphi}^\mathrm{g}_i(\theta)$ and
\eqref{eqn:marg:lik:gaussian:approx}, $-185.7$; and by using $\hat{\varphi}^\mathrm{k}_i(\theta)$ and numerical integration of the left-hand side of
\eqref{eqn:marg:lik:kernel:approx}, $-163.2$.

We have used $p=1$ for this example so that the likelihood is available, enabling
comparison with MCMC and calculation of the true marginal likelihood. However, we stress
that PW-ABC can be easily generalised for $p>1$, a case for which the likelihood is
essentially intractable and therefore one has to resort to either exact but less direct
methods (such the \linebreak Expectation$-$Maximization (EM) algorithm or data-augmented
MCMC, both of which involve treating the terms $\alpha_i \circ X_{t-i}$ and $Z_t$ as
missing data) or methods of approximate inference, such as conditional least squares which
involves minimizing $\sum_t(X_t - E[X_t|X_{t-1}])^2$; see, for example, \citet{McKen03}
and references therein.

\subsection{Stochastic Lotka--Volterra model} 
\label{sec:appl-lotka-volt}

The stochastic Lotka--Volterra (LV) model is a model of predator--prey dynamics and an
example of a stochastic discrete-state-space continuous-time Markov process \cite[see, for
example,][]{Wil06}.
Predator--prey dynamics can be thought of in chemical kinetics terms: the
predators and prey are two populations of ``reactants'' subject to three ``reactions'',
namely prey birth, predation and predator death.  Exact simulation of such models is
straightforward, e.g., using the algorithm of Gillespie (1977).  Inference is simple if
the type and precise time of each reaction is observed.  However, a more common setting is
where the population sizes are only observed at discrete time points. In this case the
number of reactions that have taken place is unknown and therefore the likelihood is not
available and hence inference is much more difficult. Reversible-jump MCMC has been developed in
this context \citep{boys_wilkinson_kirkwood:bays_inf_sto_models:08} but it requires
substantial expertise and input from the user to implement.  Particle MCMC (pMCMC) methods
\citep{Andrieu_PMCMC}, which provide an approximation to the likelihood via a Sequential
Monte Carlo (SMC) algorithm within an MCMC algorithm, have recently been proposed for
stochastic chemical kinetics models \citep{Golightly+Wilkinson:Bayes_inf_using_pMCMC:11}.
Although being computationally intensive, such methods can work reliably provided the
process is observed with measurement error. 
The {\tt R} package {\tt smfsb}, which accompanies \cite{Wil06}, contains a
pMCMC implementation designed for stochastic chemical kinetics models, and we use this
package to compare results for PW-ABC and pMCMC for the following example.

% The stochastic Lotka--Volterra (LV) model is a model of predator--prey dynamics and an
% example of a stochastic discrete-state-space continuous-time Markov process \cite[see, for
% example,][]{Wil06}.  Models of this type commonly arise when modelling chemical kinetics.
% Predator--prey dynamics can be thought of in chemical kinetics terms: the predators and
% prey are two populations of ``reactants'' subject to three ``reactions'', namely prey
% birth, predation and predator death. Inference is simple if the type and precise time of
% each reaction is observed.  However, a more common setting is where the population sizes
% are only observed at discrete time points, in which case the likelihood is not available
% and inference is much more difficult. Reversible-jump MCMC and particle MCMC methodology
% has been developed in this context \citep{boys_wilkinson_kirkwood:bays_inf_sto_models:08,
% Golightly+Wilkinson:Bayes_inf_using_pMCMC:11} but require substantial 
% expertise to implement as well as a considerable amount of input from the user. On the other
% hand, because simulating realisations from such models can be performed straightforwardly
% (for instance by using the algorithm of \cite{Gil77}) simulation-based approaches are an
% attractive alternative.

Let $Y_1$ and $Y_2$ denote the number of prey and predators
respectively, and suppose $Y_1$ and $Y_2$ are subject to the following reactions
\begin{equation}
 Y_1       \overset{r_1}{\rightarrow} 2\,Y_1, \qquad
 Y_1 + Y_2 \overset{r_2}{\rightarrow} 2\,Y_2, \qquad
 Y_2       \overset{r_3}{\rightarrow} \emptyset,
 \label{eqn:LV:reactions}
\end{equation}
which respectively represent prey birth, predation and predator death.  We consider the
problem of making inference about the rates $(r_1, r_2, r_3)$ based on
observations of $Y_1$ and $Y_2$ made at fixed intervals.

%
%% Old LV example
%
% \begin{figure}[!t] \centering \includegraphics[angle=0, width=8cm]{Figure_LV_Process} %-90 for eps 
%   \caption{A realisation of the Lotka--Volterra process generated
%     using $\mathbf{r}=(10,0.01,10)$ with $\mathbf{Y}(0)=(1000,1000)$, and
%   and the sparse dataset of 15 observations (cirles) used for inference.  Within the 
%   14 observed intervals there are 32203 events (prey births, interactions and predator
%   deaths) in the true underlying process. (Colour online)}
%  \label{fig:TLV_obs}
% \end{figure}
%
% \begin{figure}[!t]
%  \centering
%  \includegraphics[angle=0, width=8cm]{Figure_LV_Postr} %-90 for eps
%  \caption{Marginal posterior densities of the transformed parameters for the
%    the Lotka--Volterra example.  Shown are the posteriors from PW-ABC using
%    $\varepsilon=0$ for the kernel density (solid) and Gaussian (dashed) approximations for
%    each $\varphi_i(\theta)$.  Grey lines are the priors, vertical red lines show the true parameter
%    values, and red lines show the ODE-approximate (particle) MCMC posterior. (Colour online)}
%  \label{fig:TLV}
% \end{figure}

\begin{figure*}[!t]
 \centering
 \includegraphics[angle=0, width=8cm]{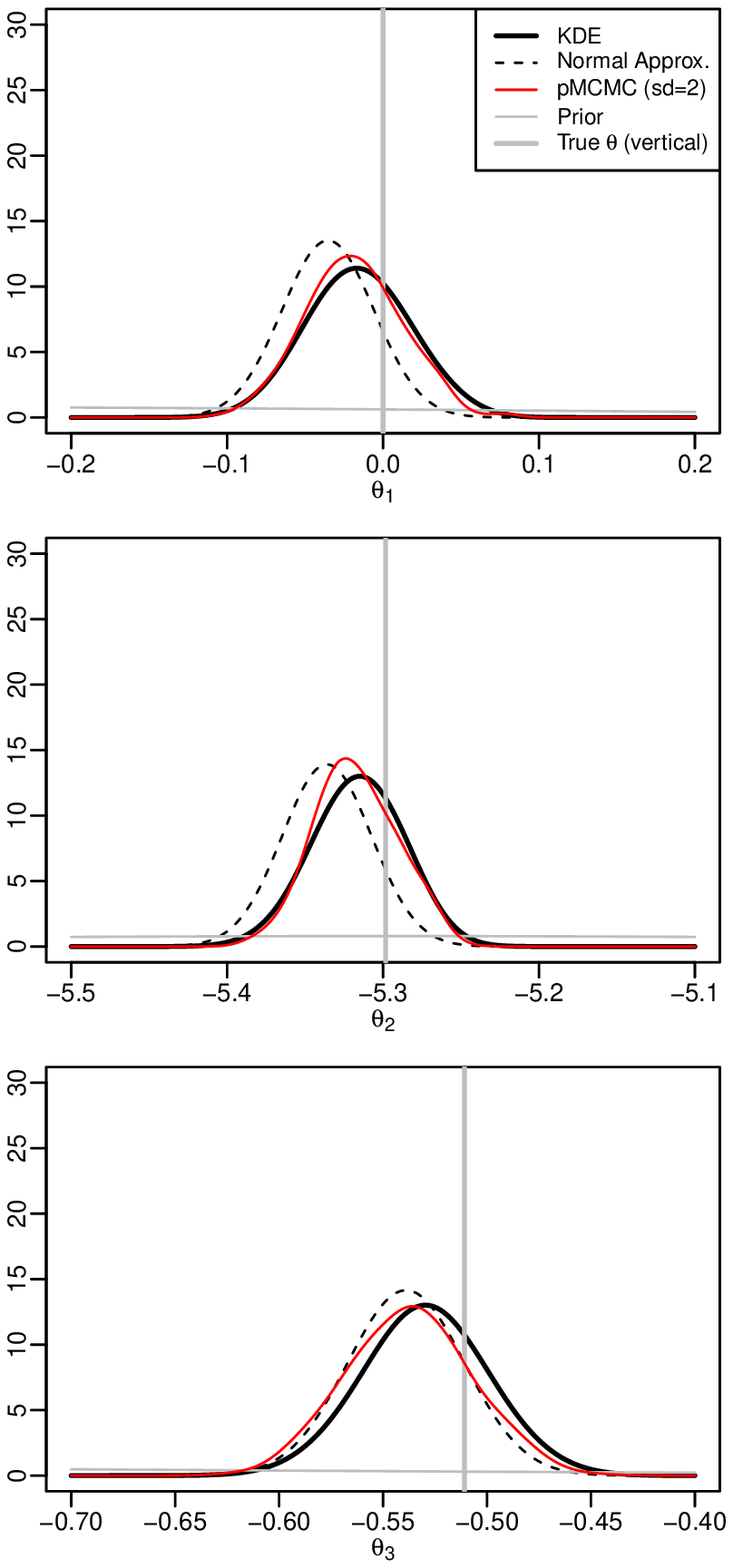} %-90 for eps
 \centering
 \includegraphics[angle=0, width=8cm]{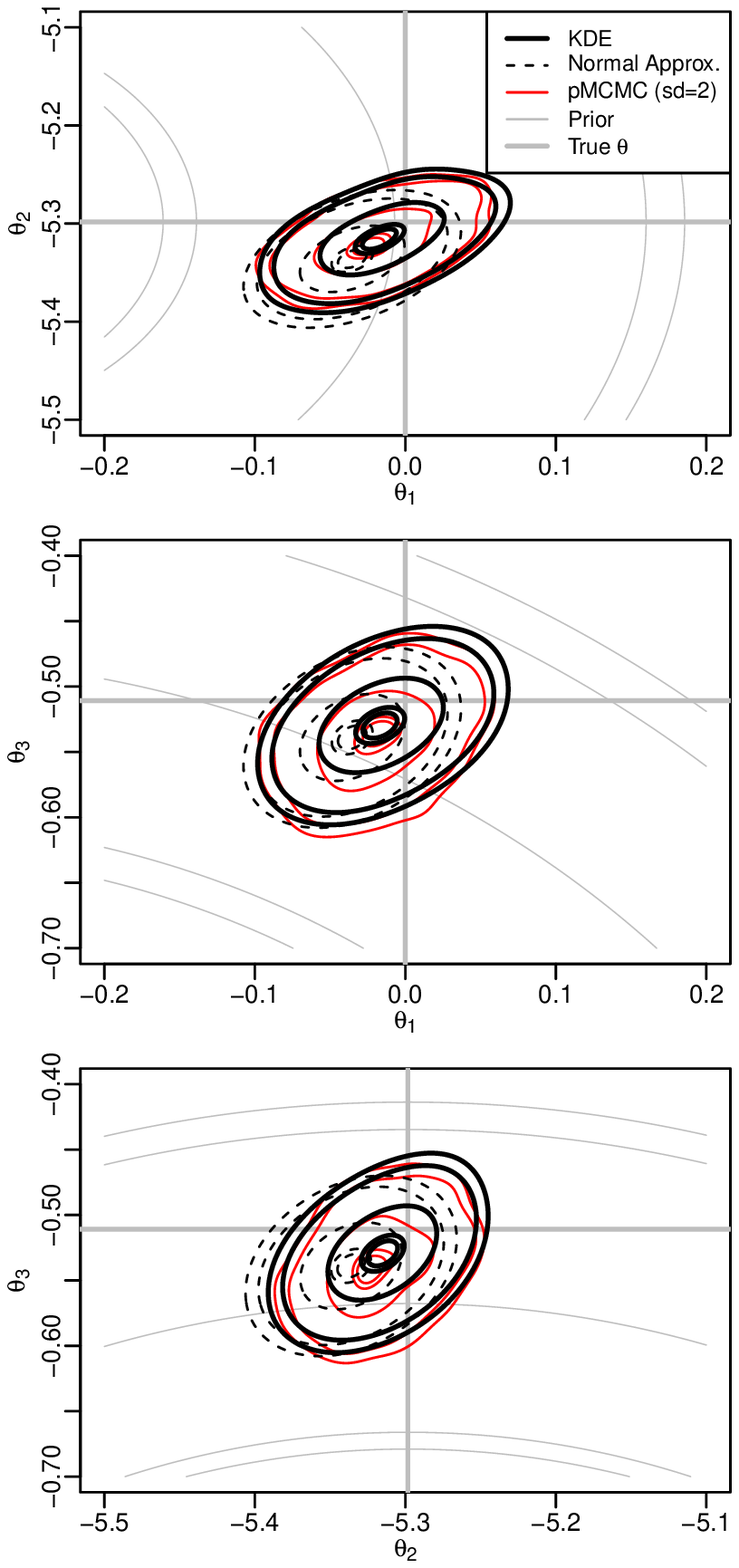} %-90 for eps
  \caption{Results for the Lotka--Volterra example
  of \S\ref{sec:appl-lotka-volt}, showing univariate and bivariate marginal posterior
  densities of ${\theta}$ based on a posterior sample from a pMCMC algorithm, and from
  the Gaussian- and kernel-based PW-ABC approximations, 
  $\hat{\pi}^\mathrm{g}(\theta|x)$ and
  $\hat{\pi}^\mathrm{k}(\theta|x)$.  For the kernel approximation we used $q=5$
  as the smoothing parameter in 
  \eqref{eqn:bandwidth}.  The contours shown in the bivariate plots are those
  that contain 5\%, 10\%, 50\%, 90\% and 95\% of probability mass.   } 
\label{fig:LV}
\end{figure*}

We generated a realisation from the stochastic LV example of \citet[][page 208]{Wil06},
that is, model \eqref{eqn:LV:reactions} using $(r_1,r_2,r_3) = (1,0.005,0.6)$, $Y_1(0)=50$
and $Y_2(0)=100$.  We performed inference in terms of transformed parameters, ${\theta} =
(\theta_1,\theta_2,\theta_3) = (\log {r}_1,\log {r}_2,\log {r}_3)$, this time with priors
$\pi(\theta_1)\sim\mathrm{Norm}( \log(0.7), 0.5 )$, $\pi(\theta_2)\sim\mathrm{Norm}(
\log(0.005), 0.5)$, and $\pi(\theta_3)\sim\mathrm{Norm}( \log(0.3), 0.5 )$.  We again
applied PW-ABC using $\varepsilon=0$, in other words requiring an exact match between the
observed and the simulated observations, to draw samples of size $m=10,000$ for each
$\varphi_i(\theta)$. { Unlike the binomial, CIR and INAR examples where drawing posterior
samples for the $\varphi_i(\theta), i=1,...,n$ assuming $\epsilon=0$ took a total of
approximately, 1, 2 and 20 minutes respectively on a standard desktop machine, for this
example doing so was computationally more demanding. However, since sampling in PW-ABC is
embarrassingly parallel we were able to draw the required samples in 32 hours on a 48
core machine.}
%, drawing $m=10,000$ samples for each $i$.   

% As a benchmark, we also performed inference using the particle MCMC
% algorithm in the {\tt smfsb} package in {\tt R}.  In initial attempts,
% we encountered problems when using a relatively small number of
% particles (in the order of 100s) mainly due to the issue that the
% maximum weight (among all particles) was negligible. Increasing the
% number of particles to 10,000 avoided the problem of the particles
% having negligible weights. However, we were unable to run the
% algorithm without having to assume an error model on the observations.

% As a benchmark, we also performed inference using the particle MCMC
% algorithm in the {\tt smfsb} package in {\tt R}.  In initial attempts,
% we found problems with the large number of particles needed to avoid
% particle degeneracy.  We found that assuming Gaussian errors (with
% $\sigma=2$) and using 10,000 particles resulted to an algorithm that
% was feasible to run in a reasonable time.

To obtain pMCMC results we found it necessary to assume an error model for the observations,
hence we assumed errors to be IID Gaussian with mean zero and standard deviation equal to
2.  Results are displayed in Figure~\ref{fig:LV}, which shows plots for univariate and
pairwise bivariate marginal posterior densities for the pMCMC results, and for the PW-ABC
approximations, $\hat{\pi}^\textrm{g}({\theta}|x)$ and
$\hat{\pi}^\textrm{k}({\theta}|x)$. Both of the PW-ABC approximations agree well
with each other and with the pMCMC results for this example.

\section{Conclusion and Discussion}

PW-ABC works by factorising the posterior density, for which
targetting by ABC would
entail a careful choice of $s(\cdot)$ and/or a large tolerance
$\varepsilon$, into a product
involving densities $\varphi_i(\theta)$, each amenable to using ABC with
$s=\textrm{Identity}(\cdot)$ and small or zero $\varepsilon$.   Having
sampled from
each $\varphi_i(\theta)$ the question then becomes how to estimate
$\pi(\theta\vert x)$
using these samples.  PW-ABC works by constructing density approximations
$\hat{\varphi}_i(\theta)$ to each $\varphi_i(\theta)$.  The approach of taking
$\hat{\varphi}_i(\theta)$ to be Gaussian, with moments matched to the
sample moments, is
computationally cheap, and if the prior is also taken to be Gaussian
then there is a
closed form expression for the Gaussian posterior density and marginal
likelihood, making
calculations extremely fast.  Taking $\hat{\varphi}_i(\theta)$ to be
Gaussian is perhaps
adequate in many applications: performance was strong in two of the
four examples we
considered.  The poor performance in the CIR and INAR examples was due
to skewness of at
least some of the $\varphi_i(\theta)$.  In the INAR example it is
striking to see an
effect so strong when the true posterior, and many of the
$\varphi_i(\theta)$, are so
close to Gaussian.  Unfortunately, increasing the number, $m$, of ABC
samples is no remedy
to this problem: as $m \rightarrow \infty$, the normalised product of
Gaussian densities,
itself Gaussian, in general does not converge to the Gaussian density
closest in the
Kullback--Leibler sense to the target density.

Two referees suggested the possibility of testing, across all of the
$\varphi_i(\theta)$,
whether a Gaussian approximation is appropriate.  A wide literature
exists on testing
multivariate normality (see \citet{Szekely200558} for a recent
contribution, plus many
references therein to earlier work) and this seems a promising
direction, but further work
is needed to devise, and understand the properties of, a procedure
based on applying
these tests in the multi-testing setting of PW-ABC.

In terms of asymptotic performance, using the kernel approximation,
$\hat{\varphi}^\textrm{k}_i(\theta)$, for $\hat{\varphi}_i(\theta)$ is
preferable since,
in this case, the estimated posterior density converges to the target
as $m \rightarrow
\infty$.  The kernel approach is computationally more demanding,
however, and its
practical use is probably limited to problems in which $\theta$ has
small dimension.  It
also requires a heuristic choice of a scalar smoothing parameter.
Choosing this parameter
too large will inflate the posterior variance; although, this aside,
in the examples we
have considered we have found posterior inference to be fairly robust
to the choice.  A
referee asked for guidance on how to choose $m$.  It is difficult to
offer general
practical advice, because the $m$ needed will depend on the dimension
of $\theta$, and on
the number and nature of the $\varphi_i(\theta)$.  The larger the
better, of course; one
possibility for checking whether $m$ is large enough might be to use a
resampling approach
to confirm that the variance, under resampling, of the
$\hat{\pi}^{\textrm k}(\theta|x)$
is acceptably small.

A possibility that generalises the Gaussian and kernel approaches, which we will
explore in future work, is to let $\hat{\varphi}_i(\theta)$ be a
mixture of, say, $u$
Gaussians (see \citet{Fan_etal:12} for an example of Gaussian mixtures
being used in a
related context).  This encompasses
\eqref{eqn:gaussian:approx:to:component:density} and
\eqref{eqn:KDE} as special cases, with $u=1$ and $u=m$ respectively.
For a general
mixture model for $\hat{\varphi}_i(\theta)$, each of the component Gaussians is
parameterised by a scalar weight, a mean vector and a covariance
matrix which need to be
determined.  We would envisage regularising, e.g., by setting each
covariance to be equal
up to scalar multiplication, perhaps as for \eqref{eqn:KDE} taking the
covariance
proportional to the sample covariance, and then fitting each
$\hat{\varphi}_i(\theta)$
based on the samples from $\varphi_i(\theta)$ using, say, an EM
algorithm.  This approach
is a compromise between \eqref{eqn:gaussian:approx:to:component:density} and
\eqref{eqn:KDE}.  It does not share the property of \eqref{eqn:KDE}
that estimated
densities converge to the true densities as $m \rightarrow \infty$,
but on the other hand
it is computationally much less involved and offers much extra freedom
and flexibility
over \eqref{eqn:gaussian:approx:to:component:density}, particularly
for dealing with
multimodal densities.  If $u$ is taken sufficiently small then it may
be feasible to work
explicitly with the $(n-1)^u$-term resulting Gaussian mixture, $\prod
\hat{\varphi}_i(\theta)$, enabling explicit calculations involving the
posterior density,
such as computing the marginal likelihood, analogous to
\eqref{eqn:marg:lik:gaussian:approx}, and direct sampling from the
approximate posterior
density (see \S\ref{sec:sampling:from:posterior}).

Several further generalisations of the PW-ABC approach are possible.  In
\eqref{eqn:factorised:likelihood}, each of the $n-1$ factors $\pi(x_i
| x_{i-1}, \theta)$,
$i=2,\ldots,n$ is the likelihood for a single data point conditional
on the previous.  An
alternative possibility is to factorise the likelihood into fewer
factors, with each
corresponding to a ``block'' of multiple observations, e.g.,
$\pi(x_{i+v_i}, x_{i+v_i-1},
\ldots , x_i | x_{i-1}, \theta)$ for some choice of $v_i$, and the
factorised likelihood
becomes a product over the relevant subset of $i=2,\ldots,n$.  To an
extent this potentially
reintroduces difficulties that with PW-ABC we sought to avoid, namely
lower acceptance rates
leading to a possible need to use a summary statistic and non-zero
tolerance (and the
ensuing ABC error they bring).  On the other hand, we might expect,
owing to the central
limit theorem, that a factor ${\varphi}_i(\theta)$ which depends on
several data points
will be closer to Gaussian than a factor dependent on only a single
data point, and hence
that \eqref{eqn:gaussian:approx:to:component:density} and
\eqref{eqn:KDE} (especially the
former) will perform better.

If using larger ``blocks'' of data in the factorisation makes it
necessary to use a
non-zero tolerance $\varepsilon>0$ (or if $\varepsilon>0$ is necessary
even when using a
single observation per factor) then there are theoretical advantages
to using what
\citet{Fearnhead+Prangle} call ``noisy ABC''.  In the context of this
paper, noisy ABC
would involve replacing the summary statistic $s(\cdot)$ with a random variable
$s'(\cdot)$ which has density uniform on a ball of radius
$\varepsilon$ around $s(\cdot)$.
Using noisy ABC ensures that, under mild regularity conditions, as $n
\rightarrow
\infty$, the posterior converges to a point mass at the true parameter
value; see \S2.2
of \citet{Fearnhead+Prangle}.

Recently, we have learnt of an interesting paper by
\citet{Barthelme+Chopin:EP-ABC:11} who
have developed an approach termed {\em Expectation Propagation-ABC}
(EP-ABC) that shares
similarities with ours.  EP-ABC is an ABC adaptation of the
Expectation Propagation
approach developed by \citet{Minka}.  EP-ABC uses essentially the same
factorisation as
\eqref{eqn:factorised:posterior} and makes a Gaussian approximation to
the density of each
factor analogous to \eqref{eqn:gaussian:approx:to:component:density}.
But then EP-ABC
proceeds rather differently: instead of drawing ABC samples for, say,
the $i$th factor by
sampling from the prior, EP-ABC draws samples from an iteratively
updated pseudo-prior.
The pseudo-prior is a Gaussian approximation to the component of the
posterior that
involves all the data \emph{except} those pertaining to the $i$th
factor.  The use of the
pseudo-prior offers a high acceptance rate in the ABC sampling and so EP-ABC can
potentially lead to an extremely fast approximation to the full
posterior $\pi(\theta |
x)$.  A disadvantage is that conditions sufficient for the convergence of EP-ABC
(or even the simpler deterministic EP) are not known.  Also, as with
using PW-ABC with
\eqref{eqn:product:of:gaussians}, since EP-ABC uses a Gaussian
approximation for each
factor, it is potentially ill-suited to problems with complicated
(e.g. multimodal or
otherwise non-Gaussian) likelihoods; convergence of the product
density is not assured to
any ``optimal'' approximation to the target posterior.  A promising
direction for
future work will be to investigate adapting the EP-ABC idea of sampling from a
pseudo-prior to the ideas in this paper of using kernel (or Gaussian
mixture) density
estimates for each likelihood factor.

\section*{Acknowledgements}
\label{sec:acknowledgements}

S.R. White was supported by the (UK) Medical Research Council [Unit Programme number
U105260794] and the EPSRC [University of Nottingham, Bridging the Gaps].  The authors
gratefully acknowledge valuable discussions with Andy Wood and Richard Wilkinson, 
and helpful comments from the anonymous referees.

\appendix

\section*{Appendix 1}

Expression for $B_{j_2,\ldots,j_n}$, $a_{j_2,\ldots,j_n}$, and $w_{j_2,\ldots,j_n}$
 in \eqref{eqn:prod:phi:kernel}, analogous to 
\eqref{eqn:B:product}--\eqref{eqn:w:product}, are as follows:
\begin{align*}
   & B_{j_2,\ldots,j_n} = \left( \sum_{i=2}^n H_i^{-1} \right)^{-1}, \\  
   &a_{j_2,\ldots,j_n} = B_{j_2,\ldots,j_n}\left( \sum_{i=2}^n H_i^{-1} {\theta}^*_{i(j_i)} \right),\\
   & w_{j_2,\ldots,j_n} = m^{(1-n)} \det(2 \pi  B_{j_2,\ldots,j_n})^{1/2} \prod_{i=2}^n \det(2 \pi
  H_i)^{-1/2} \times \nonumber \\
  &\quad \quad
  \prod_{s=2}^n 
  \prod_{t>s}^n
  \exp\left(
  -\tfrac{1}{2}{ ({\theta}^*_{s(j_s)} - {\theta}^*_{t(j_t)})^T R_{st}
  ({\theta}^*_{s(j_s)} - {\theta}^*_{t(j_t)} }
  \right), \\
  & R_{st} = H_s^{-1} B_{j_2,\ldots,j_n} H_t^{-1}.
\end{align*}
Expressions for $B'_{j_2,\ldots,j_n}$, $a'_{j_2,\ldots,j_n}$, and 
$w'_{j_2,\ldots,j_n}$ in \eqref{eqn:post:kernel} 
%are
% \begin{align*}
%  B'_{j_2,\ldots,j_n} = & \, \left( (2-n) \Sigma_\text{pri}^{-1} + B_{j_2,\ldots,j_n}^{-1}
% \right)^{-1},\\
%  a'_{j_2,\ldots,j_n} = & \,B'_{j_2,\ldots,j_n} \left( (2-n) \Sigma_\text{pri}^{-1} \,
% \mu_{\text{pri}} + B_{j_2,\ldots,j_n}^{-1} a_{j_2,\ldots,j_n} \right), \\
%  w'_{j_2,\ldots,j_n} = & \, w_{j_2,\ldots,j_n} (\det B'_{j_2,\ldots,j_n})^{1/2}
% (\det B_{j_2,\ldots,j_n})^{-1/2}  \\
% & \times (\det (2 \pi \Sigma_\text{pri}))^{(n/2-1)}
% \end{align*}
are given respectively by the right-hand sides of \eqref{eqn:post:Sigma},
\eqref{eqn:post:mu}, and \eqref{eqn:marg:lik:gaussian:approx} with 
$B$ replaced by $B_{j_2,\ldots,j_n}$, $a$ replaced by $a_{j_2,\ldots,j_n}$, and 
$w$ replaced by $w_{j_2,\ldots,j_n}$.

\section*{Appendix 2}

Proposition 1. {\em Let $I=\mathds{1}\{\theta^*\mbox{ is accepted}\}$ be the indicator
function of whether an ABC draw $\theta^*$ is accepted.
The acceptance rate is $$p(I=1) = V \,
\tilde{\pi}_\text{ABC}(x)$$ where $\tilde{\pi}_{\text{ABC}}(x)$ is the marginal likelihood
of the implied ABC posterior.} \vspace{0.2in}

\noindent {\bf Proof}

\noindent Recall from Section \ref{sec:PWABC:posterior} that $ K_{\varepsilon,p}(z) =
V^{-1} \mathds{1} \{\| z \|_p \leq \varepsilon \}$ and $\tilde{\pi}_{\text{ABC}}(x \vert
\theta) = \int \pi(y \vert \theta) K_{\varepsilon,p} \left( y - x \right) {\mathrm d} y$
is the implied ABC likelihood approximation. Then

\begin{eqnarray*}
\mathbbm{P}(I=1) & = & \int_{\theta} \mathds(I=1, \theta) \mbox{ d}\theta \\
       & = & \int_{\theta} \pi(\theta) \mathbbm{P} (I=1|\theta) \mbox{ d}\theta \\
       & = & \int_{\theta} \pi(\theta) \left\{\int_{y} \pi(y|\theta)\mathds{1}\{||y-x||_p
       \leq \varepsilon\} )\mbox{ d}y \right\} \mbox{ d}\theta  \\
       & = & \int_{\theta} \pi(\theta) \left\{\int_{y} \pi(y|\theta) V K_{\varepsilon, p}(y-x)\mbox{ d}y \right\} \mbox{ d}\theta \\
       & = & \int_{\theta} V \pi(\theta) \tilde{\pi}_{ABC} (x|\theta)   \mbox{ d}\theta \\
       & = & V \, \tilde{\pi}_\text{ABC}(x).
\end{eqnarray*}
\vspace{-0.5in}
\begin{flushright}
 $\square$
\end{flushright}

\bibliographystyle{plainnat}

\begin{thebibliography}{30}
\providecommand{\natexlab}[1]{#1}
\providecommand{\url}[1]{\texttt{#1}}
\expandafter\ifx\csname urlstyle\endcsname\relax
  \providecommand{\doi}[1]{doi: #1}\else
  \providecommand{\doi}{doi: \begingroup \urlstyle{rm}\Url}\fi

\bibitem[Al-Osh and Alzaid(1987)]{Al-OsAl87}
M.A. Al-Osh and A.A. Alzaid.
\newblock First-order integer-valued autoregressive ({INAR}({$1$})) process.
\newblock \emph{J. Time Ser. Anal.}, 8\penalty0 (3):\penalty0 261--275, 1987.

\bibitem[Andrieu et~al.(2010)Andrieu, Doucet, and Holenstein]{Andrieu_PMCMC}
C.~Andrieu, A.~Doucet, and R.~Holenstein.
\newblock Particle {M}arkov chain {M}onte {C}arlo methods.
\newblock \emph{J. R. Stat. Soc. Ser. B Stat. Methodol.}, 72\penalty0
  (3):\penalty0 269--342, 2010.

\bibitem[{Barthelm{\'e}} and {Chopin}(2011)]{Barthelme+Chopin:EP-ABC:11}
S.~{Barthelm{\'e}} and N.~{Chopin}.
\newblock {Expectation-Propagation for Summary-Less, Likelihood-Free
  Inference}.
\newblock \emph{ArXiv e-prints}, 2011.

\bibitem[Beaumont et~al.(2002)Beaumont, Zhang, and
  Balding]{beaumont_zhang_balding:abc_pop_genetics:02}
M.A. Beaumont, W.~Zhang, and D.J. Balding.
\newblock {A}pproximate {B}ayesian {C}omputation in {P}opulation {G}enetics.
\newblock \emph{Genetics}, 162\penalty0 (4):\penalty0 2025--2035, December
  2002.

\bibitem[Blum and Fran\c{c}ois(2010)]{BlumFran10}
M.G.B. Blum and O.~Fran\c{c}ois.
\newblock Non-linear regression models for approximate bayesian computation.
\newblock \emph{Statistics and Computing}, 20:\penalty0 63--73, 2010.

\bibitem[Boys et~al.(2008)Boys, Wilkinson, and
  Kirkwood]{boys_wilkinson_kirkwood:bays_inf_sto_models:08}
R.J. Boys, D.J. Wilkinson, and T.B. Kirkwood.
\newblock Bayesian inference for a discretely observed stochastic kinetic
  model.
\newblock \emph{Stat. Comput.}, 18\penalty0 (2):\penalty0 125--135, 2008.

\bibitem[Cox et~al.(1985)Cox, Ingersoll, and Ross]{MR785475}
J.C. Cox, J.E. Ingersoll, and S.A. Ross.
\newblock A theory of the term structure of interest rates.
\newblock \emph{Econometrica}, 53\penalty0 (2):\penalty0 385--407, 1985.

\bibitem[{Dean} et~al.(2011){Dean}, {Singh}, {Jasra}, and {Peters}]{Peters2010}
T.~A. {Dean}, S.~S. {Singh}, A.~{Jasra}, and G.~W. {Peters}.
\newblock {Parameter Estimation for Hidden Markov Models with Intractable
  Likelihoods}.
\newblock \emph{ArXiv e-prints}, 2011.

\bibitem[Fan et~al.(2012)Fan, Nott, and Sisson]{Fan_etal:12}
Y.~Fan, D.J. Nott, and S.A. Sisson.
\newblock {Approximate Bayesian computation via regression density estimation}.
\newblock \emph{Technical report, arXiv:1212.1479}, 2012.

\bibitem[Fearnhead and Prangle(2012)]{Fearnhead+Prangle}
P.~Fearnhead and D.~Prangle.
\newblock {Constructing summary statistics for approximate Bayesian
  computation: semi-automatic approximate Bayesian computation}.
\newblock \emph{RSS Series B}, 2012.
\newblock In Press.

\bibitem[Fukunaga(1972)]{fukunaga}
K.~Fukunaga.
\newblock \emph{Introduction to statistical pattern recognition}.
\newblock Electrical science series. Academic Press, 1972.

\bibitem[Gabriel et~al.(2010)Gabriel, Wilson, Leatherbarrow, Cheesbrough, Gee,
  Bolton, Fox, Fearnhead, Hart, and
  Diggle]{Wilson_Gabriel_Leatherbarrow_Cheesbrough_Gee_Bolton_Fox_Hart_Diggle_%
Fearnhead_2009}
E.~Gabriel, D.J. Wilson, A.J. Leatherbarrow, J.~Cheesbrough, S.~Gee, E.~Bolton,
  A.~Fox, P.~Fearnhead, C.A. Hart, and P.J. Diggle.
\newblock Spatio-temporal epidemiology of campylobacter jejuni enteritis, in an
  area of northwest england, 2000--2002.
\newblock \emph{Epidemiology and Infection}, 138:\penalty0 1384--1390, 2010.

\bibitem[Golightly and
  Wilkinson(2011)]{Golightly+Wilkinson:Bayes_inf_using_pMCMC:11}
A.~Golightly and D.J. Wilkinson.
\newblock Bayesian parameter inference for stochastic biochemical network
  models using particle markov chain monte carlo.
\newblock \emph{Interface Focus}, 1\penalty0 (6):\penalty0 807--820, 2011.

\bibitem[Mardia et~al.(1979)Mardia, Kent, and Bibby]{Mardia_etal79}
Kantilal~Varichand Mardia, John~T. Kent, and John~M. Bibby.
\newblock \emph{Multivariate analysis}.
\newblock Academic Press [Harcourt Brace Jovanovich Publishers], London, 1979.
\newblock ISBN 0-12-471250-9.
\newblock Probability and Mathematical Statistics: A Series of Monographs and
  Textbooks.

\bibitem[Marin et~al.(2012)Marin, Pudlo, Robert, and Ryder]{Marin2011}
J.M. Marin, P.~Pudlo, C.~Robert, and R.~Ryder.
\newblock {Approximate Bayesian computational methods}.
\newblock \emph{Stat. Comput.}, 22\penalty0 (5):\penalty0 1009--1020, 2012.

\bibitem[Marjoram et~al.(2003)Marjoram, Molitor, Plagnol, and
  Tavar{\'e}]{MarMolPlagTav03}
P.~Marjoram, J.~Molitor, V.~Plagnol, and S.~Tavar{\'e}.
\newblock {Markov chain Monte Carlo without likelihoods}.
\newblock \emph{Proc. Natl Acad. Sci. USA}, 100\penalty0 (26):\penalty0 15324,
  2003.

\bibitem[McKenzie(2003)]{McKen03}
E.~McKenzie.
\newblock Discrete variate time series.
\newblock In \emph{Stochastic processes: modelling and simulation}, volume~21
  of \emph{Handbook of Statist.}, pages 573--606. North-Holland, Amsterdam,
  2003.

\bibitem[McKinley et~al.(2009)McKinley, Cook, and
  Deardon]{McKinley+Cook+Deardon:infer_epidemic_models_wo_likelihoods:09}
T.~McKinley, A.~Cook, and R.~Deardon.
\newblock Inference in epidemic models without likelihoods.
\newblock \emph{Intl. J. Biostat.}, 5:\penalty0 24, 2009.

\bibitem[Minka(2001)]{Minka}
T.P. Minka.
\newblock Expectation propagation for approximate bayesian inference.
\newblock In \emph{Proceedings of the 17th Conference in Uncertainty in
  Artificial Intelligence}, UAI '01, pages 362--369, San Francisco, CA, USA,
  2001. Morgan Kaufmann Publishers Inc.
\newblock ISBN 1-55860-800-1.
\newblock URL \url{http://dl.acm.org/citation.cfm?id=647235.720257}.

\bibitem[Moriña et~al.(2011)Moriña, Puig, Ríos, Vilella, and
  Trilla]{Morina+et_al:model_hospital_admissions:11}
D.~Moriña, P.~Puig, J.~Ríos, A.~Vilella, and A.~Trilla.
\newblock A statistical model for hospital admissions caused by seasonal
  diseases.
\newblock \emph{Statistics in Medicine}, 30\penalty0 (26):\penalty0 3125--3136,
  2011.

\bibitem[Neal and Subba~Rao(2007)]{NealRao07}
P.~Neal and T.~Subba~Rao.
\newblock M{CMC} for integer-valued {ARMA} processes.
\newblock \emph{J. Time Ser. Anal.}, 28\penalty0 (1):\penalty0 92--110, 2007.

\bibitem[Pritchard et~al.(1999)Pritchard, Seielstad, Perez-Lezaun, and
  Feldman]{pritchard:population_growth_of_Y_chormosomes:99}
J.K. Pritchard, M.T. Seielstad, A.~Perez-Lezaun, and M.W. Feldman.
\newblock {Population growth of human Y chromosomes: a study of Y chromosome
  microsatellites}.
\newblock \emph{Mol. Biol. Evol.}, 16\penalty0 (12):\penalty0 1791--1798, 1999.

\bibitem[Silverman(1986)]{silverman:density_estimation:86}
B.W. Silverman.
\newblock \emph{{Density Estimation for Statistics and Data Analysis}}.
\newblock {Chapman \& Hall/CRC}, London - New York, 1986.

\bibitem[Sudderth et~al.(2003)Sudderth, Ihler, Freeman, and Willsky]{ihler}
E.B. Sudderth, A.T. Ihler, W.T. Freeman, and A.S. Willsky.
\newblock Nonparametric belief propagation.
\newblock \emph{Computer Vision and Pattern Recognition, IEEE Computer Society
  Conference on}, 1:\penalty0 605, 2003.

\bibitem[Sz\'{e}kely and Rizzo(2005)]{Szekely200558}
G.J. Sz\'{e}kely and M.L. Rizzo.
\newblock A new test for multivariate normality.
\newblock \emph{Journal of Multivariate Analysis}, 93\penalty0 (1):\penalty0
  58--80, 2005.

\bibitem[Toni et~al.(2009)Toni, Welch, Strelkowa, Ipsen, and
  Stumpf]{toni_welch_strelkowa_ipsen_stumpf:abc_dynamical_systems:09}
T.~Toni, D.~Welch, N.~Strelkowa, A.~Ipsen, and M.P. Stumpf.
\newblock {Approximate Bayesian computation scheme for parameter inference and
  model selection in dynamical systems}.
\newblock \emph{J. R. Soc. Interface}, 6:\penalty0 187--202, 2009.

\bibitem[Van~Kampen(2007)]{van2007stochastic}
N.G. Van~Kampen.
\newblock \emph{Stochastic Processes in Physics and Chemistry}.
\newblock North-Holland Personal Library. Elsevier, 2007.

\bibitem[Wand and Jones(1995)]{Wand+Jones:kernel_smoothing:95}
P.~Wand and C.~Jones.
\newblock \emph{Kernel Smoothing}.
\newblock Monographs on Statistics and Applied Probability. Chapman \& Hall,
  1995.

\bibitem[Wilkinson(2011)]{Wil06}
D.J. Wilkinson.
\newblock \emph{Stochastic modelling for systems biology}.
\newblock Chapman \& Hall/CRC, 2011.
\newblock 2nd Edition.

\bibitem[Wilkinson(2013)]{Wil13}
R.D. Wilkinson.
\newblock {Approximate Bayesian computation (ABC) gives exact results under the
  assumption of model error}.
\newblock \emph{Stat. Appl. Genet. Mo. B.}, (to appear; available online as
  arXiv:0811.3355, 2009), 2013.

\end{thebibliography}

\end{document}